\documentclass[twocolumn]{aa}
\usepackage[utf8]{inputenc}
\usepackage{natbib}
\usepackage{graphicx}
\usepackage{placeins}
\usepackage{color}
\usepackage{amsmath}
\usepackage{natbib}
\usepackage{arydshln}

\newcommand{\FIG}[1]{#1}
\usepackage{geometry}
\title{Effects of ambipolar diffusion on waves in the solar chromosphere}
\author{B. Popescu Braileanu
          \inst{1}
          \and
          R. Keppens \inst{1}}
\titlerunning{Ambipolar diffusion}
\authorrunning{Popescu Braileanu and Keppens}
\institute{Centre for mathematical Plasma Astrophysics, KU Leuven, 3001 Leuven, Belgium,\email{\\beatriceannemone.popescubraileanu@kuleuven.be} }
\date{}

\date{Received XXXX; Accepted XXXX}
 
\abstract{
{\it Context.} 
The chromosphere is a partially ionized layer of the solar atmosphere, the transition between the photosphere where the gas motion is determined by the gas pressure
and the corona dominated by the magnetic field.
\\{\it Aims.}
We study the effect of partial ionization for 2D wave propagation in a gravitationally stratified, magnetized atmosphere with properties similar to the solar chromosphere. 
\\{\it Methods.}
We adopt 
an oblique uniform magnetic field in the plane of propagation with strength suitable for  a quiet sun region. 
The theoretical model used is a single fluid magnetohydrodynamic approximation, where ion-neutral interaction is modeled by the ambipolar diffusion term. Magnetic energy can be converted into internal energy through the dissipation
of the electric current produced by the drift between ions and neutrals.
We use numerical simulations where we continuously drive fast waves at the bottom of the atmosphere. The collisional coupling between ions and neutrals decreases with the decrease of the density and the ambipolar effect becomes important.
\\{\it Results.}
Fast waves excited at the base of the atmosphere reach the equipartition layer and reflect or transmit as slow waves.
While the waves propagate through the atmosphere and the density drops, the waves steepen into shocks. 
\\{\it Conclusions.}
The main effect of ambipolar diffusion is damping of the waves. 
We find that for the parameters chosen
in this work, the ambipolar diffusion affects the fast wave before it is reflected, with damping being more pronounced for waves which are launched
in a direction perpendicular to the magnetic field. Slow waves are less affected by ambipolar effects.
The damping increases for shorter periods and larger magnetic field strengths.
Small scales produced by the 
nonlinear effects and the superposition of different types of waves created at the equipartition height are efficiently damped by ambipolar diffusion.
}

\begin{document}

\maketitle

\section{Introduction}

Ambipolar diffusion is the dissipation of the magnetic field through the electric field created by the drift velocity
between neutrals and charged particles.
It is the most important process of magnetic field decay in neutron stars \citep{i1,i2,i3}. 
Ambipolar diffusion has significant effects on turbulence present
in the interstellar medium \citep{turb1} or as produced by the magnetorotational instability in weakly ionized disks \citep{mri}.
In solar context, ambipolar diffusion has an important contribution in the chromosphere.
The chromosphere is the layer of the solar atmosphere which makes the transition between the almost neutral
photosphere to the fully ionized corona. As the density drops with height, the neutrals and the charged particles
become incompletely coupled by collisions in the chromosphere.

Waves are ubiquitous in the solar atmosphere and have been observed at different layers from the photosphere to the corona.
Because of the limitation on the spatial and temporal resolution of current telescopes, only the low frequency waves could be observed so far.
However, theoretical computations of \cite{mus} for the acoustic flux generated by the solar convection showed a peak located at the high frequency waves,
corresponding to periods of $\approx$ 10 s. These are the waves we will study in this paper.

In a single fluid magnetohydrodynamic (MHD) approach, the interaction between neutrals and charges is assumed to happen on a timescale much shorter than the hydrodynamic timescale.  The effect
of ion-neutral interactions can be introduced as a contribution to the electric field in a generalized Ohm's law.
This is different from a full two-fluid model, where there are separate equations which describe the time evolution of neutrals and charges, so that 
there is no assumption
about their collisional timescale.   
It has been shown that in many situations, the single fluid MHD (extended with ambipolar terms) and the full two-fluid treatment
give very similar results \citep{turb1,mythesis}.
%In some other cases, such as slow sausage waves, the different behaviour between charged particle waves where the gravitational stratification
%imposes a cutoff for large frequency and the neutral waves where the cutoff is for small frequencies could only be captured by a fully two-fluid model \citep{bl1}.  

%The damping of the Alfv\'en waves in the solar chromosphere due to Cowling resisitivity has been studied \cite{leake1}.
In solar context, mode conversion from fast to Alfv\'en waves has been studied by taking into account the effect of  ambipolar diffusion \citep{wave2}.
This fast to Alfv\'en mode conversion happens for large periods.
% and the ambipolar diffusion does not have a considerable effect, however the small scales
%can be produced by external factor such as turbulence. 
It has been shown that this mode conversion is not significantly modified by ambipolar diffusion,
but only indirectly affected through the damping of the fast waves. The overall conclusion is that the main effect of ambipolar diffusion is damping of waves \citep{wave2}. In this paper, we will study a purely 2D configuration where the Alfv\'en waves are omitted, but look in detail to slow and fast mode propagation as affected by ambipolar diffusion.
%In some cases, the ambipolar diffusion is not related to diffusivity, as it introduces a force proportional to the magnetic field and can lead to sharp structures,
%especially around a null point \citep{sharpAmbi}.

Waves excited by convection motions in the photosphere propagate upwards, and they split into fast and slow components at the equipartition layer where
the sound speed is similar to the Alfv\'en speed. The fast wave most probably
reflects before reaching the upper part of the atmosphere \citep{wave1,wave2}.
The fast-slow conversion mechanism was proposed in order to explain absorption of acoustic waves in sunspots \citep{bog1}. The mode transformation is complete for waves propagating along the magnetic field lines \citep{Paul2}.  
Depending on the parameters of the atmosphere and the inclination angles,
the converted slow wave can propagate upwards or downwards \citep{wave1}. For a detailed review of wave propagation in sunspots see \cite{khrev}. We will look instead to quiet sun magnetic field strengths.

To study waves in a stratified, magnetized atmosphere, many linear studies are available in the literature, especially in ideal, single fluid linear MHD \citep[see e.g. Chapter 7 in ][]{hans2019}. 
These give analytical solutions for the perturbations in all variables, and offer a quicker possibility to analyze a large range of parameters compared to numerical simulations. 
%However, there are very few cases where even an approximate analytical solution is hard to find.
In this study, we will derive and use a local dispersion relation, as obtained from 
manipulating the governing linearized equations in MHD, with and without ambipolar effects. In our Appendix~\ref{appendixA}, we clarify the link between our local dispersion relation and previous linear studies. The local dispersion relation we derive here is used to implement a realistic driver in our numerical simulations at the bottom of our atmosphere.
%The simplified first-order correction calculated as in \cite{Popescu2019} is several orders of magnitude smaller than the value of the vertical wavenumber.

When nonlinear effects come into play, there are no known analytical solutions
even in simple cases, and numerical simulations should be used.
In this study we perform linear to nonlinear simulations of wave propagation in a 2D geometry using the {\tt MPI-AMRVAC} code \citep[][\url{http://amrvac.org/}]{amrvac}. 
We first describe the implementation of the ambipolar term in the code, and give details about the setup and the simulations performed in 
section \ref{sec2}.
We present results of simulations mainly in the linear regime in section \ref{sec:res}.
Section \ref{sec:nonlin} shows a more complex setup of nonlinear interaction of two pulses. 
We analyze the heating associated with the propagation of the waves in Section \ref{sec:heat}.
The discussion and conclusions are summarized in section \ref{sec:cl}.

%%%%%%%%%%%%%%%%%%%%%%%%%%%%%%%%%%%%%%%%%%%%%%%%%%%%%%%%%%%%%%%%%%%%%%%%%%%%%%%%%%%%%%%%%%%%%%%%%%%%%%%%%%%%
\section{Description of the problem} \label{sec2}
%%%%%%%%%%%%%%%%%%%%%%%%%%%%%%%%%%%%%%%%%%%%%%%%%%%%%%%%%%%%%%%%%%%%%%%%%%%%%%%%%%%%%%%%%%%%%%%%%%%%%%%%%%%%

\subsection{Governing nonlinear equations}\label{ss:eqns}

The single fluid, nonlinear MHD equations, extended with the ambipolar effect, can be written in terms of the density $\rho$, velocity $\mathbf{u}$, internal energy density $e_{\rm int}$ and magnetic and electric fields $\mathbf{B}$ and $\mathbf{E}$ as follows:
\begin{eqnarray} \label{eq:eqs_mhd_ambi}
\frac{\partial \rho}{\partial t} + \mathbf{\nabla}\cdot \left(\rho\mathbf{u}\right) =  0\,, \nonumber  \\ 
\frac{\partial (\rho\mathbf{u})}{\partial t} + \mathbf{\nabla}\cdot (\rho\mathbf{u} \mathbf{u} +p)  = \mathbf{J}\times\mathbf{B} + \rho\mathbf{g}\,, \nonumber \\
\frac{\partial e_{\rm int}}{\partial t}  +  \nabla \cdot ( \mathbf{u} e_{\rm int}  )  = -p \nabla \cdot \mathbf{u} +
\mathbf{J} \cdot \mathbf{E}_{\rm nonideal} \,, \nonumber\\ 
\frac{\partial\mathbf{B}}{\partial t}  =  -\mathbf{\nabla}\times \mathbf{E}\,.
\end{eqnarray}
These express mass conservation, the momentum equation with on the right hand side the Lorentz force (using the current density $\mathbf{J}=\nabla\times\mathbf{B}$) and external gravity (with acceleration $\mathbf{g}$), and they include a deviation from ideal MHD by ambipolar (charged-neutral decoupling) effects in the generalized Ohm's law:
\begin{equation} 
\mathbf{E} =- \mathbf{u} \times \mathbf{B} + \mathbf{E}_{\rm nonideal}\,,
\end{equation}
where the ambipolar term is generically given by
\begin{equation} 
\mathbf{E}_{\rm nonideal} =  -\eta_A [(\mathbf{J} \times \mathbf{B}) \times \mathbf{B}] \,.
\end{equation}
Writing the current density as $\mathbf{J}=\left(\mathbf{J}\cdot\mathbf{B}\right)\mathbf{B}/B^2+\mathbf{J}_{\perp B}$ we observe that this perpendicular current density is
\begin{equation} 
\mathbf{J}_{\perp B} =  -\frac{[(\mathbf{J} \times \mathbf{B}) \times \mathbf{B}]}{B^2} \,,
\end{equation}
so the heating due to ambipolar effects is given by
\begin{equation} 
 \mathbf{J} \cdot \mathbf{E}_{\rm nonideal} = \eta_A {J}_{\perp B}^2 B^2.
\end{equation}
This acts as a local source term for the internal energy density $e_{\rm int}=p/(\gamma-1)$, which closes the equations and introduces the ratio of specific heats $\gamma$.

In order to obtain an expression for the ambipolar coefficient $\eta_A$ in the single-fluid model, further assumptions about the neutral and charged fluid temperatures must be used, 
and the ionization fraction has to be estimated, e.g. from the Saha equation. When doing so,
the ambipolar diffusivity coefficient can be defined as in \cite{eqkh},
\begin{equation} \label{eq:etaa-total}
\eta_A=\frac{\xi_n^2}{\alpha \rho_n \rho_c},
\end{equation}
where $\xi_n=\rho_n/\rho$ denotes the neutral fraction since $\rho_n$ gives the neutral density as opposed to the charged density $\rho_c$, with $\rho=\rho_n+\rho_c$. This Eq.~(\ref{eq:etaa-total}) involves a non-trivial collisional parameter $\alpha$ that in general depends on the collisional effects incorporated { and the ionization fraction}. In this work, we will rather simplify the actual 
functional dependence on the density to its essential ${1}/{\rho^2}$, so that the ambipolar coefficient becomes
\begin{equation} \label{eq:eff_alpha}
\eta_A = \frac{\nu_A}{\rho^2}, 
\end{equation}
with $\nu_A$ a constant, input parameter.
%as defined as:
%\begin{equation} \label{eq:eff_alpha}
%\alpha = \frac{m_{\rm in}}{{m_{\rm n}}^2} \sqrt{ \frac{8 k_{\rm B} T_{\rm cn}}{\pi m_{\rm in}}} 
 %\sigma^{\rm elastic}_{\rm in} + \frac{m_{\rm en}}{{m_{\rm n}}^2} \sqrt{ \frac{8 k_{\rm B} T_{\rm cn}}{\pi m_{\rm en}}}  \sigma^{\rm elastic}_{\rm en}\,.
%\end{equation}
%\noindent

%%%%%%%%%%%%%%%%%%%%%%%%%%%%%%%%%%%%%%%%%%%%%%%%%%%%%%%%%%%%%%%
\subsection{2D equilibrium setup}\label{ss:setup}
%%%%%%%%%%%%%%%%%%%%%%%%%%%%%%%%%%%%%%%%%%%%%%%%%%%%%%%%%%%%%%%
\begin{figure}[!htb]
\centering
\FIG{\includegraphics[width=8cm]{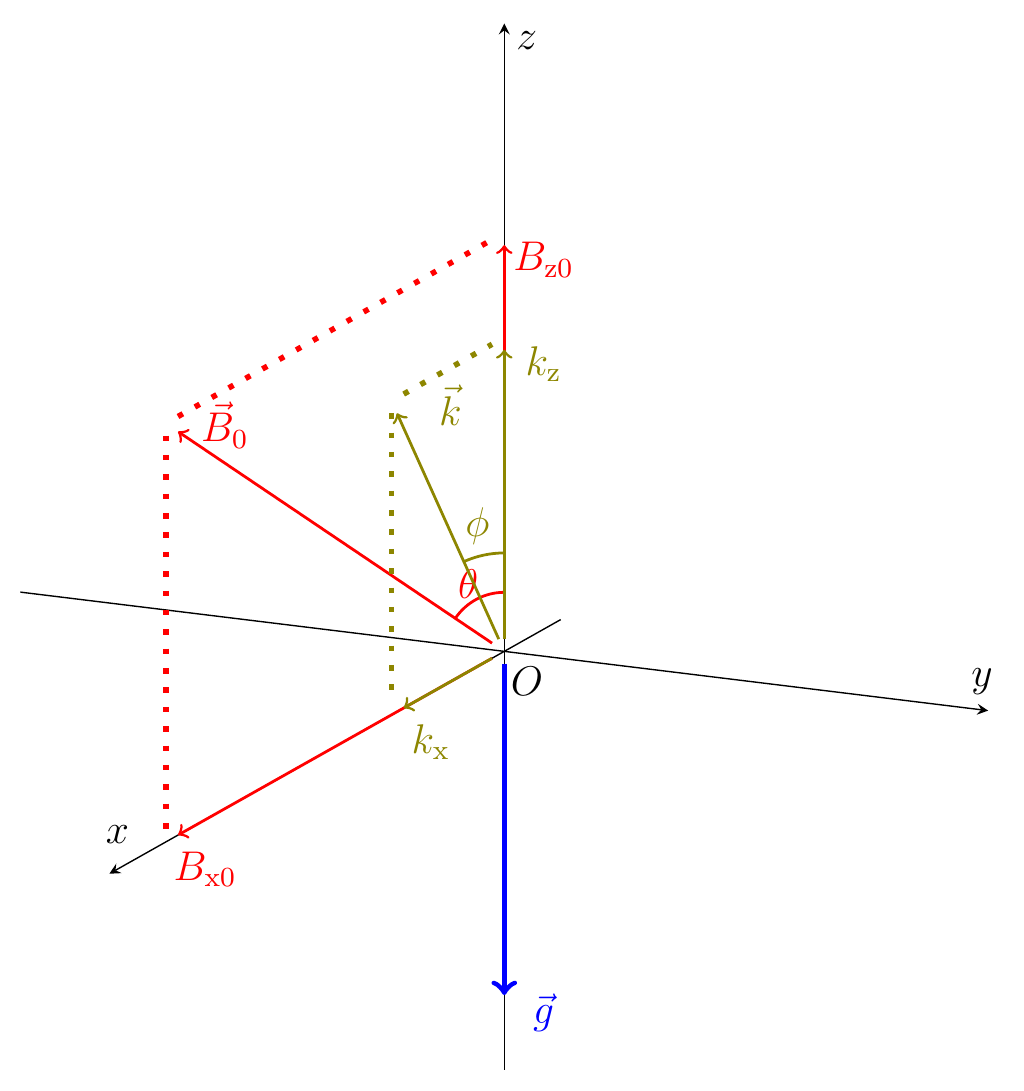}}
\caption{Sketch of the configuration, with gravity along $-z$, and wavevector $\mathbf{k}$ and magnetic field $\mathbf{B}$ in the $x-z$ plane.
}
\label{fig:setup1}
\end{figure}
%%%%%%%%%%%%%%%%%%%%%%%%%%%%%%%%%%%%%%%%%%%%%%%%%%%%%%%%%%%%%%%
\begin{figure*}[!htb]
\centering
\FIG{\includegraphics[width=8cm]{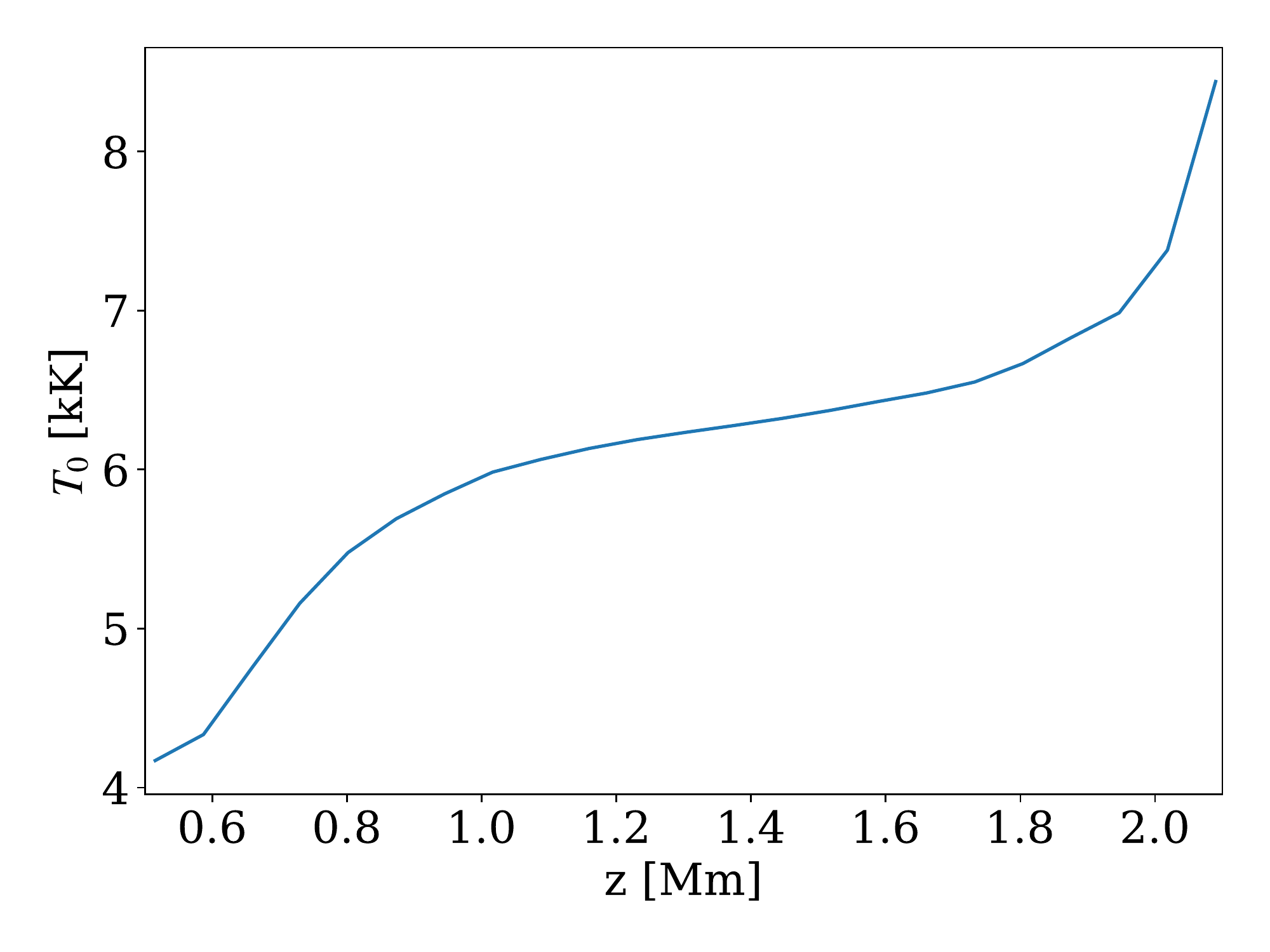}
\includegraphics[width=8cm]{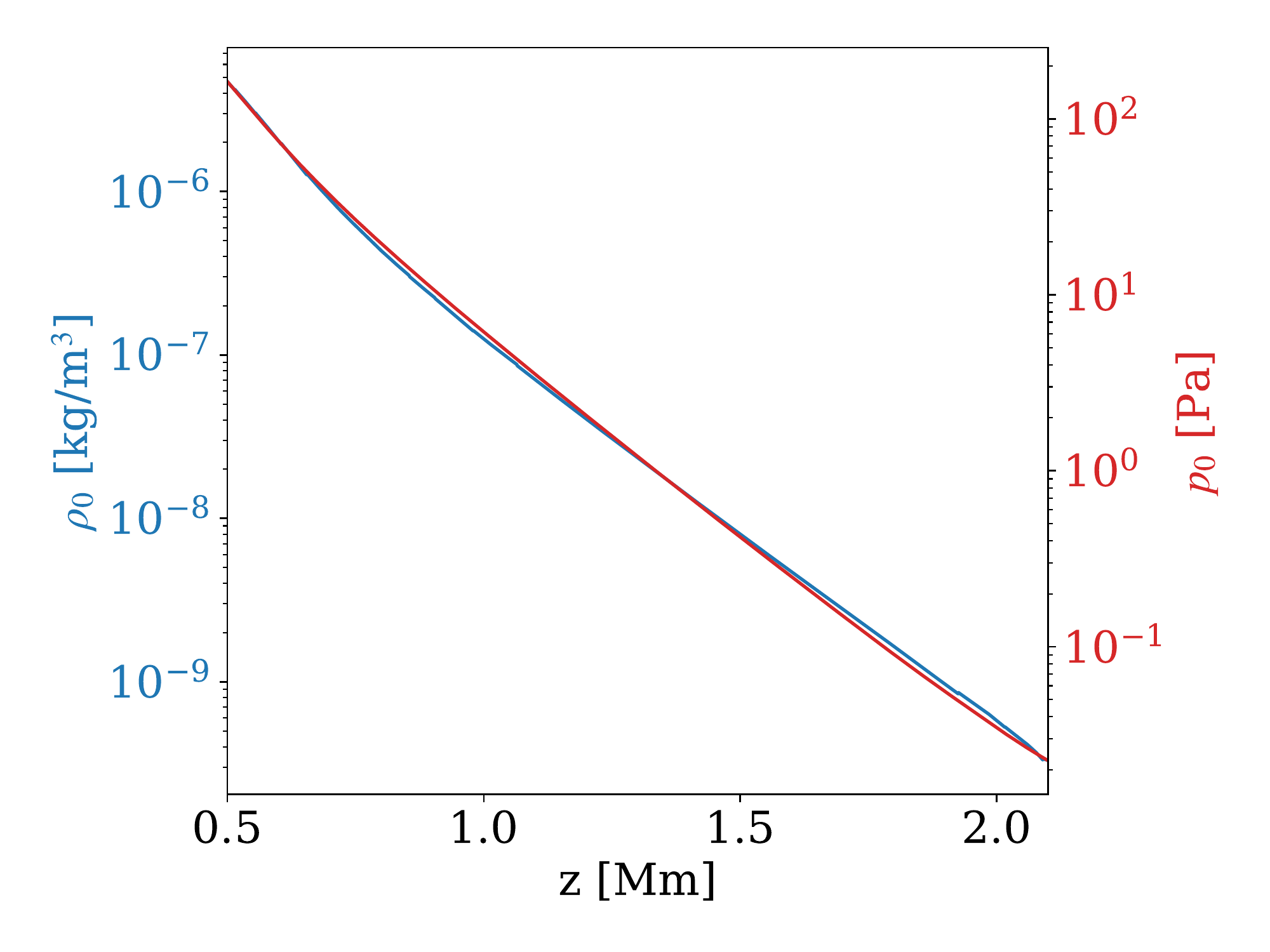}}
\caption{The equilibrium temperature (left) and density-pressure (right) as a function of height $z$.
}
\label{fig:setup}
\end{figure*}
%%%%%%%%%%%%%%%%%%%%%%%%%%%%%%%%%%%%%%%%%%%%%%%%%%%%%%%%%%%%%%%
\begin{figure}[!htb]
\centering
\FIG{\includegraphics[width=8cm]{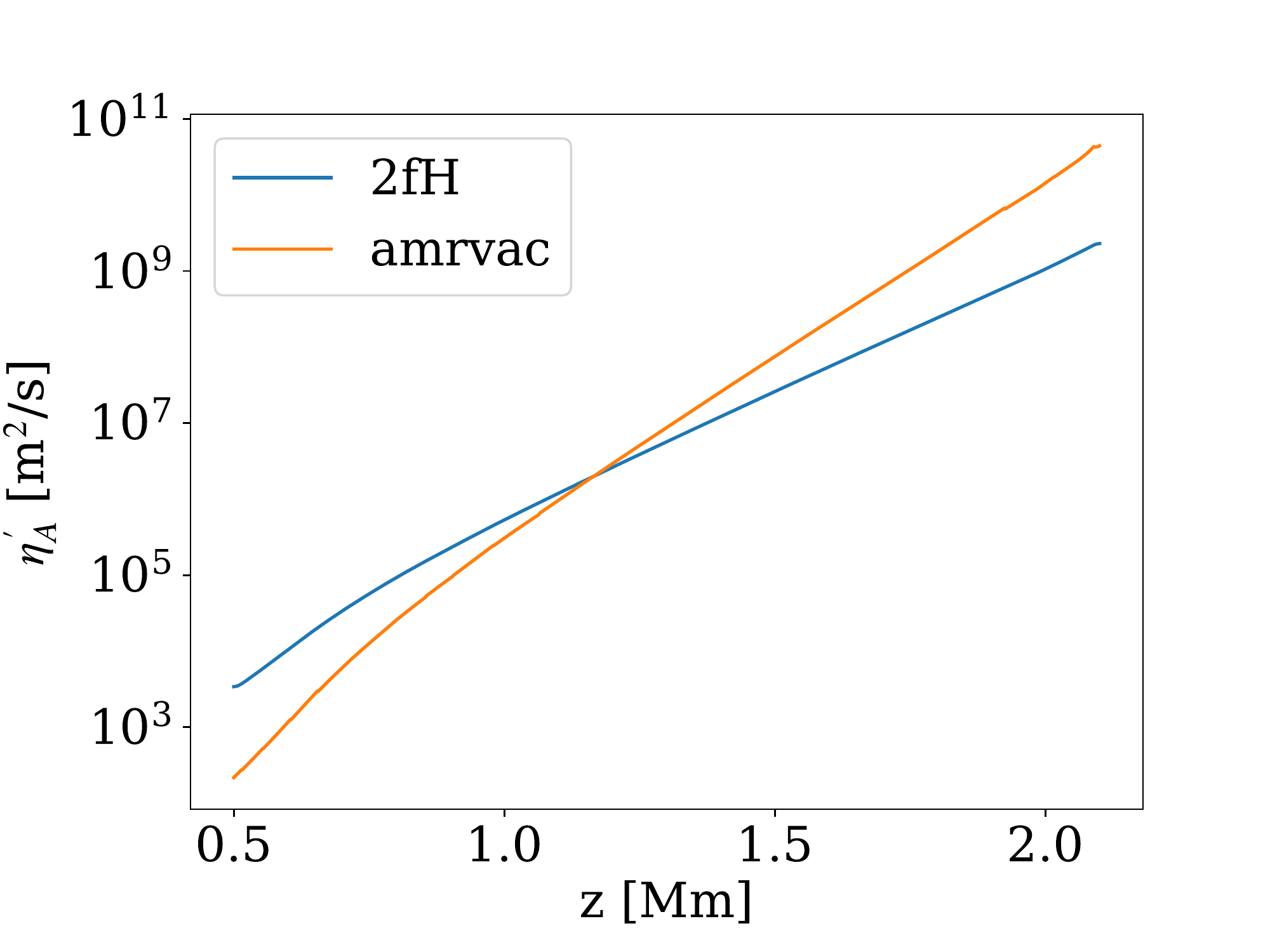}}
\caption{
Ambipolar coefficient in units of m$^2$/s as calculated from a two fluid model (blue line), and the coefficient used in the simulations (orange line)
}
\label{fig:nuA}
\end{figure}
%%%%%%%%%%%%%%%%%%%%%%%%%%%%%%%%%%%%%%%%%%%%%%%%%%%%%%%%%%%%%%%
\begin{table*}
\caption{List of the simulations used in the paper. The columns represent, from left to right:
the abbreviated name of the simulation used in the text, and the parameters of the simulations.}
\begin{tabular}{llllllll}
\hline
Name & Type of the perturbation &$L_x$ [Mm]&  $B_0$ [G] & $\theta$ [$^\circ$]& Period [s]& $\phi$ [$^\circ$] & A\\
\hline
P-s  & plane& 1&17.4 & 25.8 & 5 & 22.3 & 10$^{-3}$\\
G-s  & gaussian ($x_c=-0.6$ Mm)& 2& 17.4 & 25.8 & 5 & 22.3 & 10$^{-3}$\\
G-sr  & gaussian ($x_c=0.6$ Mm)& 2& 17.4 & 25.8 & 5 & -22.3 & 10$^{-3}$\\
P  & plane& 1& 17.4 & 25.8 & 5 & 10.9 & 10$^{-3}$ \\
G  & gaussian ($x_c=-0.6$ Mm)& 2& 17.4 & 45 & 5 & 10.9 & 10$^{-3}$\\
P-f  & plane& 1& 17.4 & 84.2 & 5 & 10.9 & 10$^{-3}$\\
P-a  & plane& 1& 17.4 & 25.8 & 5 & 49.5 & 10$^{-3}$\\
P-as  & plane& 1& 17.4 & 49.5 & 5 & 49.5 & 10$^{-3}$\\
P-p  & plane& 1& 17.4 & 25.8 & 20 & 49.5 & 10$^{-3}$\\
P-pf  & plane& 1& 17.4 & 84.2 & 20 & 49.5 & 10$^{-3}$\\
P-b  & plane& 1 & 52.3 & 25.8 & 5 & 10.9 & 10$^{-3}$\\
2G  & two gaussians ($x_c=-0.6$; $0.6$ Mm)& 3& 17.4 & 25.8 & 5 & 22.3; -22.3 & 10$^{-1}$; 10$^{-1}$\\
%2G,res2  & \multicolumn{6}{l}{same as 2G, twice resolution in both $x$ and $z$ directions}\\
\hline
\end{tabular}
\label{tab:sims}
\end{table*}
%%%%%%%%%%%%%%%%%%%%%%%%%%%%%%%%%%%%%%%%%%%%%%%%%%%%%%%%%%%%%%%

We study the propagation of waves in a 2D geometry illustrated in Figure~\ref{fig:setup1}, where we will model the $x-z$ plane.
We use a gravitationally stratified atmosphere in the $z$-direction with a uniform magnetic field contained in the $x-z$ plane.
The equilibrium atmosphere is invariant along the $x$-direction. 
In the MHD approach, this equilibrium with pressure $p_0(z)$ and density $\rho_0(z)$ fulfills the hydrostatic differential equation
\begin{equation}
\frac{d p_0}{d z} = -\rho_0 g\,,
\end{equation}
where $g$ is the magnitude of the gravitational acceleration $\mathbf{g}$, assumed to be uniform.
We use the temperature profile from the VALC model \citep{VALC} and adopt the number density at the base of the atmosphere $n_0=2.84 \times 10^{21}$ m$^{-3}$.
Figure \ref{fig:setup} shows the equilibrium profiles of the temperature (left panel), density and pressure (right panel).
The magnetic field strength used in most simulations is $B_0 = 17.4$ G, which makes the gas pressure equal to the magnetic pressure at the middle of our vertical domain, which extends from between 0.5 Mm and 2.1 Mm in height. 
{ The equipartition layer defined as the height where the sound speed equals the Alfv\'en speed is located higher than the height where
the magnetic pressure equals the gas pressure, at $\approx$~0.697 of the vertical domain.}
The equipartition layer  separates the region of weak field regime at the bottom
of the atmosphere from the region of strong field regime in the upper part of the atmosphere.
When we use instead $B_0 = 52.3$ G, { the magnetic pressure equals the gas pressure at 
a height located at 1/4 of the vertical domain size. 
}
 
In order to obtain a  value for $\nu_A$ suitable for the VALC  model atmosphere used in this work,
we averaged the actual $\eta_A \rho^2$ in height from the full two-fluid model used in \cite{Popescu2019}, which gave a value of 
$\nu_A = 3.17 \times 10^{-9}$ s kg/m$^3$.
The code {\tt MPI-AMRVAC} \citep[][\url{http://amrvac.org/}]{amrvac}  uses nondimensional variables and the simulations presented in this work use a normalization for lengths equal to 
$10^{6} \text{ m}$, for temperature measured in $5000 \text{ K}$ and for number density $10^{20} \text{ m}^{-3}$ and adopts magnetic units where the magnetic permeability is unity. In addition, we considered a plasma composed of Hydrogen only, and we used  a ratio of specific heats $\gamma=5/3$. This normalization gave a value of $\nu_A \approx 10^{-4}$ in nondimensional units, which is used in all our simulations.

{
The  value of the ambipolar coefficient in units of m$^2$/s, $\eta_A'=\eta_A/\mu_0$, calculated from the two-fluid model, composed by Hydrogen only, gravitationally stratified \citep{Popescu2019}
and the value used in this work, for a magnetic field with strength $B_0 = 17.4$ G, are shown for comparison in  Figure \ref{fig:nuA} \citep[see also Fig. 5 from][]{eqkh}.
The ambipolar coefficient used in our simulations has a steeper increase with height, the difference being caused mainly by keeping a fixed  ionization fraction in our case.
We neglected other non-ideal effects such as Hall effect and Ohmic dissipation which have a larger contribution at the bottom of the atmosphere
than the ambipolar effect \citep[Fig. 5][]{eqkh}, because we wanted to distinguish the ambipolar effect.  The ambipolar effect has a large contribution towards the upper boundary,
where the collision frequency between ions and neutrals decreases.
}
%
%%%%%%%%%%%%%%%%%%%%%%%%%%%%%%%%%%%%%%%%%%%%%%%%
\subsection{Wave driving}

The vertical domain between 0.5 Mm and 2.1 Mm is always covered by 3200 grid points. We generate a specific type of wave perturbation in the ghostcells at the bottom of our atmosphere, and this wave starts propagating in 
the unperturbed atmosphere described above. The horizontal domain varies with the type of wave perturbation that we study: it is between -0.5 Mm and 0.5 Mm with 400 points for plane waves,
between -1 Mm and 1 Mm with 600 points for the single gaussian waves, and
between -1.5 Mm and 1.5 Mm with 800 points for the two gaussian wave simulations. The horizontal domain length is also specified in Table \ref{tab:sims}.
We always use periodic boundary conditions in the $x$-direction and open boundary conditions for the perturbed quantities in the vertical direction for the upper boundary.
{ The equations evolved by the code use a splitting of the density, pressure and magnetic field variables
into time-independent and time-dependent parts, where the time-independent variables correspond to the equilibrium variables $\rho_0$, $p_0$ and $\mathbf{B_0}$, mentioned above.
In order to avoid reflection at the upper boundary we used an absorbing layer of 0.1 Mm thickness at the top of our atmosphere. This layer is uniform, the  gravity  also being considered zero there.
The time-dependent variables are damped exponentially in this layer:
\begin{equation}
u' = u \cdot \text{exp}\left( - \sigma_d \frac{z-z_{\rm db}}{z_{\rm dt} - z_{\rm db}} \right)\,,
\end{equation}
where $z_{\rm db} = 2.1$ Mm and $z_{\rm dt} = 2.2$ Mm are the bottom and the top of the layer respectively, and $\sigma_d$ is a damping coefficient, constant for each simulation.
}

We used a three step numerical scheme for the temporal integration, 
and the HLL  method  for the calculation of the fluxes \citep{hll}
with a third order slope limiter \citep{cada3}. As the  ambipolar parabolic term imposes a very restrictive explicit CFL timestep,
it has been implemented using the supertimestepping (STS) method as described in the two variants available in the literature,
one which uses Legendre polynomials, also known as RKL
 \citep{sts1-ref,sts1-ref1,sts1-ref2,sts1-ref4,sts1-ref3}, 
and the other which uses Chebyshev polynomials, also known as RKC \citep{sts2-ref,sts2-ref3,sts2-ref4,sts2-ref2,sts2-ref1}.
For the simulations presented in this work we used the RKL method and a splitting technique of each temporal substep which adds  the ambipolar contribution 
before and after the contribution of the rest of the terms using half of the substep timestep.
This splitting  makes the temporal scheme second order accurate.

Our wave driver is based on an approximate analytical solution as explained in the next section~\ref{ss:solution}. 
This involves (1) an amplitude for the imposed vertical velocity $V_z = A c_0$, where $A$ is a free parameter and  $c_0=\sqrt{\gamma p_0/\rho_0}$ is the (local) sound speed, (2)
the period of the wave, and (3) the horizontal wave number $k_x$. We then
obtain the vertical wave number $k_z$ by solving a dispersion relation given by Eq.~(\ref{eq:disp}). From the four solutions we will choose the one corresponding to the fast upward wave.
We will use two different shapes for the perturbation: a plane wave and a gaussian wave, which uses a gaussian function described by:
\begin{equation}\label{eq:gaussf}
g(x) = \text{exp}\left( -\frac{(x-x_c)^2}{2 \sigma^2} \right)\,.
\end{equation}
We used $\sigma=8$ km for all the simulations that have gaussian perturbation and $x_c$ varied as shown in Table \ref{tab:sims}.
There are five parameters which are varied in our simulations: the amplitude parameter $A$, the angle of the magnetic field with the vertical direction ($\theta$), the strength of the initial uniform magnetic field $B_0$,
the period of the wave and the local value of the angle $\phi$ between the propagation vector and the vertical direction at the base of the atmosphere.
This angle $\phi$ is calculated from $k_x$ and $k_z$, with $k_z$ obtained from the dispersion relation.
The simulation parameters used in this paper are summarized in Table \ref{tab:sims}.

%%%%%%%%%%%%%%%%%%%%%%%%%%%%%%%%%%%%%%%%%%%%%%%%%%%%%%%%%%%%%%%%%%%%%%%%%%%%%%%%%%%%
\subsection{Approximate analytical solution}\label{ss:solution}
%%%%%%%%%%%%%%%%%%%%%%%%%%%%%%%%%%%%%%%%%%%%%%%%%%%%%%%%%%%%%%%%%%%%%%%%%%%%%%%%%%%%

%%%%%%%%%%%%%%%%%%%%%%%%%%%%%%%%%%%%%%%%%%%%%%%%%%%%%%%%%%%%%%%%%%%%%%%%%%%%%%%%%%%%

For the waves, we linearize the MHD equations given by Eq.~(\ref{eq:eqs_mhd_ambi}) about the static equilibrium atmosphere. These linearized MHD equations, in terms of perturbations $\rho_1$, $p_1$, velocities $v_x$, $v_y$, and $v_z$ and $\mathbf{B}_1$, for our 2D geometry with its uniform magnetic field, taking into account the ambipolar diffusivity are 
\begin{align} \label{eqs_mhd} 
&\frac{\partial \rho_1}{\partial t} = - v_z \frac{d \rho_0}{d z} - \rho_0  \left[\frac{\partial v_x}{\partial x}+\frac{\partial v_z}{\partial z}\right]\,, \nonumber   \\
&\rho_0 \frac{\partial v_x}{\partial t} =  - \frac{\partial p_1}{\partial x} 
 + B_{\rm z0} J_{y1} \,,  \nonumber \\ 
&\rho_0 \frac{\partial v_z}{\partial t} = -\rho_1 g - \frac{\partial p_1}{\partial z} 
- B_{\rm x0} J_{y1}\,,   \nonumber \\  
&\frac{\partial p_1}{\partial t} = c_0^2 \frac{\partial \rho_1}{\partial t} + c_0^2 v_z \frac{d \rho_0}{d z} - v_z \frac{d p_0}{d z} , \nonumber\\
&\frac{\partial B_{\rm x1}}{\partial t} =  B_{\rm z0}\frac{ \partial v_x }{\partial z}  - B_{\rm x0}\frac{ \partial v_z }{\partial z} +B_0^2  \left(\frac{d \eta_{A0}}{d z}+\eta_{A0} \frac{\partial }{\partial z}\right) J_{y1}, \nonumber \\
&\frac{\partial B_{\rm z1}}{\partial t} = 
- B_{\rm z0} \frac{\partial v_x}{\partial x} + B_{\rm x0} \frac{\partial v_z}{\partial x}  
-B_0^2 \eta_{A0} \frac{\partial }{\partial x} J_{y1}\,,
\end{align}
where we used the $y$-component of the perturbed current density
\begin{align}\label{eq:jy1}
& J_{y1}=\left[\frac{ \partial B_{\rm x1} }{\partial z} -\frac{ \partial B_{\rm z1} }{\partial x}  \right]\,.
\end{align}
We also used that when $\partial/\partial y=0$, the linearized equations decouple completely such that only the in-plane velocity perturbations $v_x$, $v_z$ and magnetic perturbations $B_{\rm x1}$ and $B_{\rm z1}$ matter. The equilibrium density profile also enters in $\eta_{A0}=\nu_A/\rho_0^2$.
We now introduce the ansatz where
the dependence in time and in the $x$-direction of the perturbations are written explicitly as $\text{exp} (i \omega t)$ and $\text{exp} (-i k_x x)$, respectively.
The expressions for the vertical dependence of the perturbation in density and pressure are then obtained directly from the linearized continuity and pressure equations, when $\omega\ne 0$:
\begin{align} \label{eq:rho_p}
&\rho_1(z) = \frac{1}{\omega} \left[ k_x \rho_0 v_x  + i \left(\rho_0 \frac{d v_z}{d z}+   \frac{d \rho_0}{d z} v_z \right) \right]\,,\nonumber \\
&p_1(z) = \frac{1}{\omega} \left[ k_x \rho_0 c_0^2    v_x    + i \left( \rho_0 c_0^2  \frac{d v_z}{d z} - \rho_0  g  v_z \right) \right]\,.
\end{align}
\noindent

To make further analytic progress, we will first neglect the ambipolar effect, i.e. take $\eta_{A0}=0$.
In that case, the linearized induction equation gives
\begin{eqnarray}
B_{\rm x1}(z) &= &\frac{i}{\omega}  \left( B_{\rm x0}  \frac{d v_z}{d z}  - B_{\rm z0}  \frac{d v_x}{d z}  \right)\,,\nonumber \\
B_{\rm z1}(z) &= &\frac{k_x}{\omega}\left( B_{\rm z0}  v_x   - B_{\rm x0} v_z \right) \,. \label{eq:mhd_b}
\end{eqnarray}
After some manipulations, where we introduce the above expressions in the equations for $x$ and $z$ momentum, we obtain a system of two coupled ODEs for $v_x(z)$ and $v_z(z)$:
\begin{eqnarray}\label{eq:mhd}
\left(B_{\rm x0}^2 + \gamma p_0\right)  \frac{d^2 v_z}{d z^2}  -\gamma\rho_0 g  \frac{d v_z}{d z}  +\left( \omega^2 \rho_0 - k_x^2 B_{\rm x0}^2  \right)   v_z   && \nonumber \\
\quad \quad   -B_{\rm x0} B_{\rm z0} \frac{d^2 v_x}{d z^2}  - i k_x \gamma p_0   \frac{d v_x}{d z} && \nonumber \\  
+ \left[k_x^2 B_{\rm x0} B_{\rm z0}  - i k_x \rho_0 g (1-\gamma) \right]  v_x = 0\,, &&\nonumber \\
B_{\rm x0}  B_{\rm z0}  \frac{d^2 v_z}{d z^2}  + i k_x \gamma p_0 \frac{d v_z}{d z}  - \left( k_x^2 B_{\rm x0} B_{\rm z0} + i k_x \rho_0 g  \right)   v_z   &&  \nonumber \\
\quad \quad   -B_{\rm z0}^2 \frac{d^2 v_x}{d z^2}  
- \left[\omega^2  \rho_0- k_x^2 (B_{\rm z0}^2 + \gamma p_0) \right]  v_x = 0\,. && 
\end{eqnarray}
We can observe that when the field has no vertical component, $B_{\rm z0} =0$,  Eqns.~(\ref{eq:mhd}) reduce to a single second order ODE for the vertical velocity,
equivalent to Eq.~(5) from \cite{Nye}. In that case where $B_{\rm z0} =0$, it is also equivalent to Eq.~(7.80) from the textbook by~\cite{hans2019}, and that form shows the MHD continuous parts of the spectrum most clearly. In this special $k_y=0$ case, the Alfv\'en continuum drops out of the problem, and only leaves the slow continuum. The Eqs.~(\ref{eq:mhd}) are equivalent to the coupled second order ODEs as Eqs. (2.9) and (2.10) from \cite{Paul1}. The left hand-side terms of the equations as derived by \cite{Paul1} describe acoustic and magnetic oscillations, and coupling terms describe the magnetic and the acoustic influence on these oscillations,  respectively.

As our interest is in waves traveling through the atmosphere rather than solving the governing system of ODEs (which needs a sufficient amount of boundary conditions), we instead derive a local dispersion relation. 
We introduce in Eq.~(\ref{eq:mhd}) a wave solution for the vertical direction, assuming 
that the vertical dependence of the perturbation is of form $\text{exp} (-i k_z z)$. In this approximation, the perturbation of the magnetic field is locally perpendicular to the propagation direction at all heights, 
\begin{equation} \label{eq:b1k}
\mathbf{B}_1 \perp \mathbf{k}\,.
\end{equation}
This local assumption gives us a relationship between the amplitude of the horizontal $v_x$ and vertical velocity, $v_z$, and a local (i.e. with space-dependent coefficients)  dispersion relation, which is a fourth order equation in $k_z$, for a given real frequency $\omega$ and wavenumber component $k_x$. The details of this Eq.~(\ref{eq:disp}) are given in Appendix~\ref{appendixA}, where our approach is compared to previous treatments for waves in the magnetized, stratified atmosphere. The very same simple local dispersion analysis can be done as well with nonzero ambipolar coefficient $\eta_A\ne 0$. This case leads to more complicated expressions, where the calculations have been performed with the {\tt Mathematica} software, but is just as in the ideal MHD case
leading to a fourth order dispersion relation in $k_z$ similar to Eq.~(\ref{eq:disp}).

We use the linear analysis from this local dispersion relation in our drivers as follows: for a given period (setting the given real $\omega$) and horizontal wavenumber $k_x$, we compute the (possibly complex) $k_z$ (and hence find $\phi$ from its real part, although we can also fix $\phi$ and compute $k_x$, $k_z$ analogously) according to the local conditions (density, pressure etc) in the bottom ghost cells below our atmosphere. We then use relations~(\ref{eq:mhd_b}), (\ref{eq:rho_p}), and relations between the velocity amplitudes from Eq.~(\ref{eq:mhd1}) in setting all perturbed quantities in accord with the selected wave solution. This involves taking the real part of the $v_z=V_z\text{exp} (i\omega t-ik_x x-ik_z z)$ prescription, and similarly for all other variables. For injected plane waves, this is adopted along the entire bottom boundary, while for gaussian cases, this prescription is multiplied with a spatial gaussian window function $g$, described in Eq. (\ref{eq:gaussf}).
As the ambipolar effect is negligible at the bottom of the atmosphere we use the same prescription for the driver in both MHD and ambipolar runs.

%%%%%%%%%%%%%%%%%%%%%%%%%%%%%%%%%%%%%%%%%%%%%%%%%%%%%%%%%%%%%%%%%%%%%%%%%%%%%%%%%%%%%%%%%%%%%%%%%%%%%%
\begin{figure*}[htb]
\centering
\FIG{\includegraphics[width=16cm]{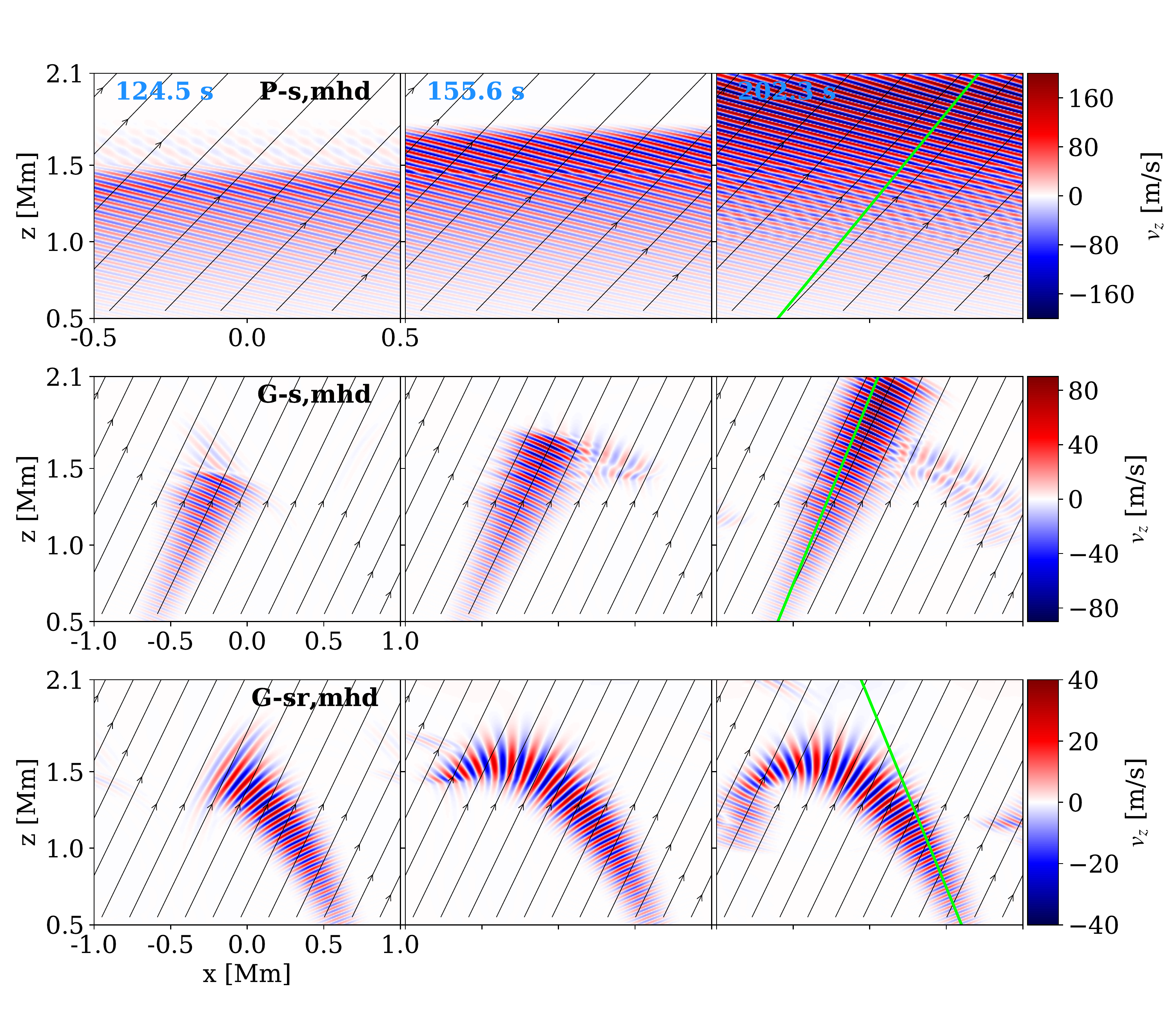}}
\caption{
Snapshots of vertical velocity for the plane wave P-s (top), the gaussian pulse G-s (middle), and the reversed gaussian G-sr (bottom) in the MHD case at three different moments. Note that the horizontal domain size differs for the three cases.
}
\label{fig:snap_g2}
\end{figure*}
%%%%%%%%%%%%%%%%%%%%%%%%%%%%%%%%%%%%%%%%%%%%%%%%%%%%%%%%%%%%%%%
%%%%%%%%%%%%%%%%%%%%%%%%%%%%%%%%%%%%%%
\begin{figure}[h]
\centering
\FIG{\includegraphics[width=8cm]{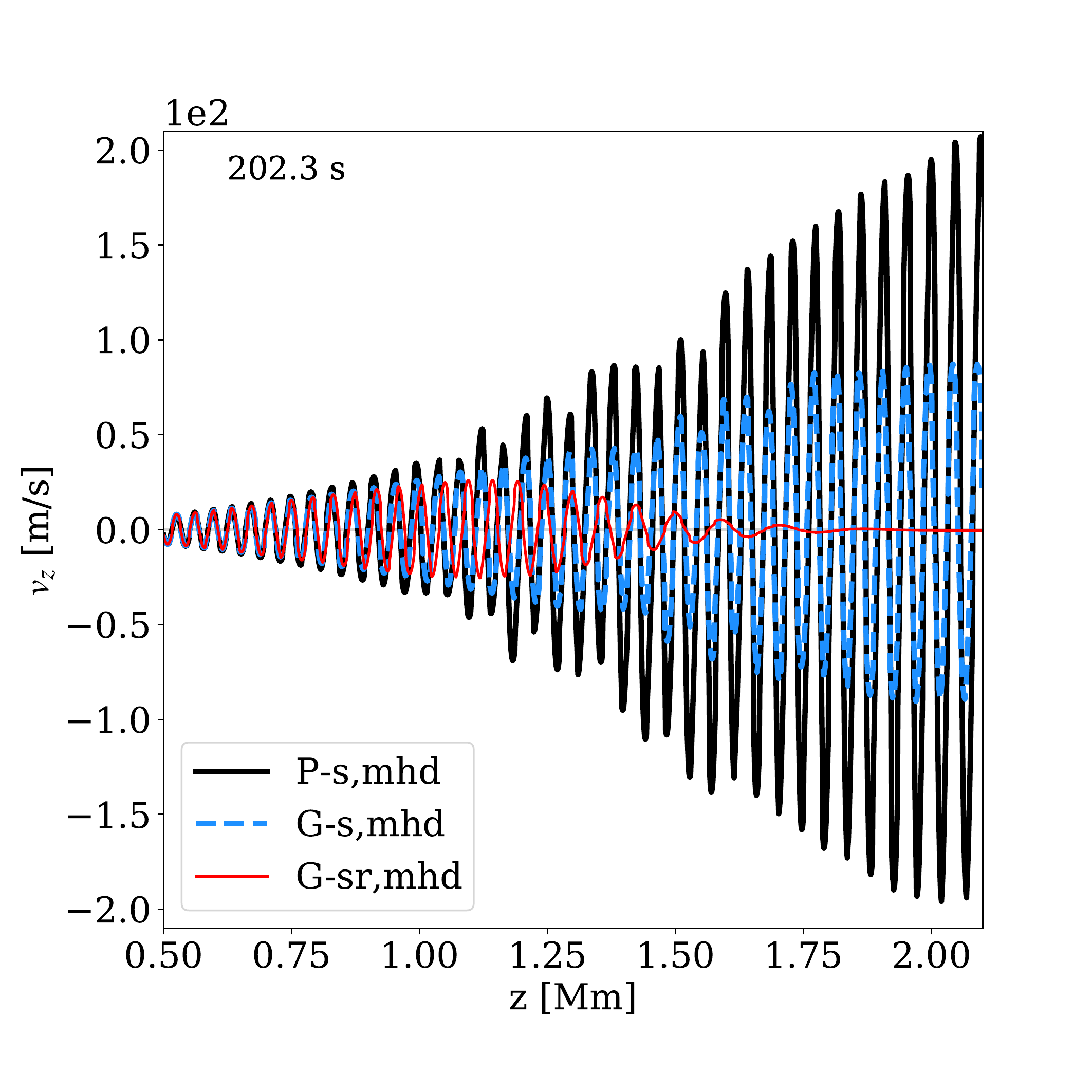}}
\caption{
%Left: 
Profile of vertical velocity along the green lines shown in Figure \ref{fig:snap_g2}, for the three cases shown there.
%Right: Profile of vertical and horizontal velocity components along $x=0$.
}
\label{fig:gs}
\end{figure}
%%%%%%%%%%%%%%%%%%%%%%%%%%%%%%%%%%%%%%%%%%%%%%%%%%%%%%%%%%%%%%%%

%%%%%%%%%%%%%%%%%%%%%%%%%%%%%%%%%%%%%%
\begin{figure*}[h]
\centering
\FIG{\includegraphics[width=8cm]{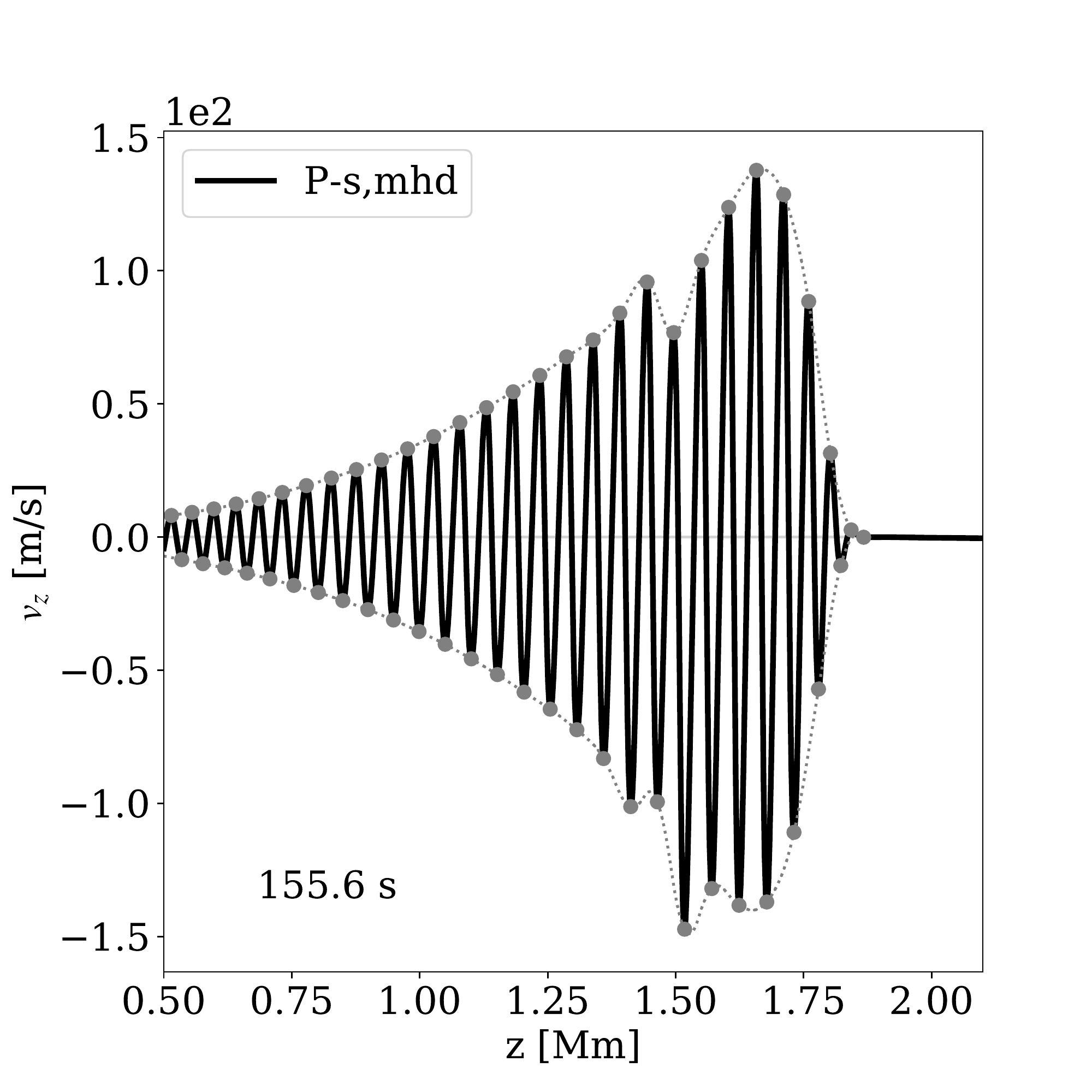}
\includegraphics[width=8cm]{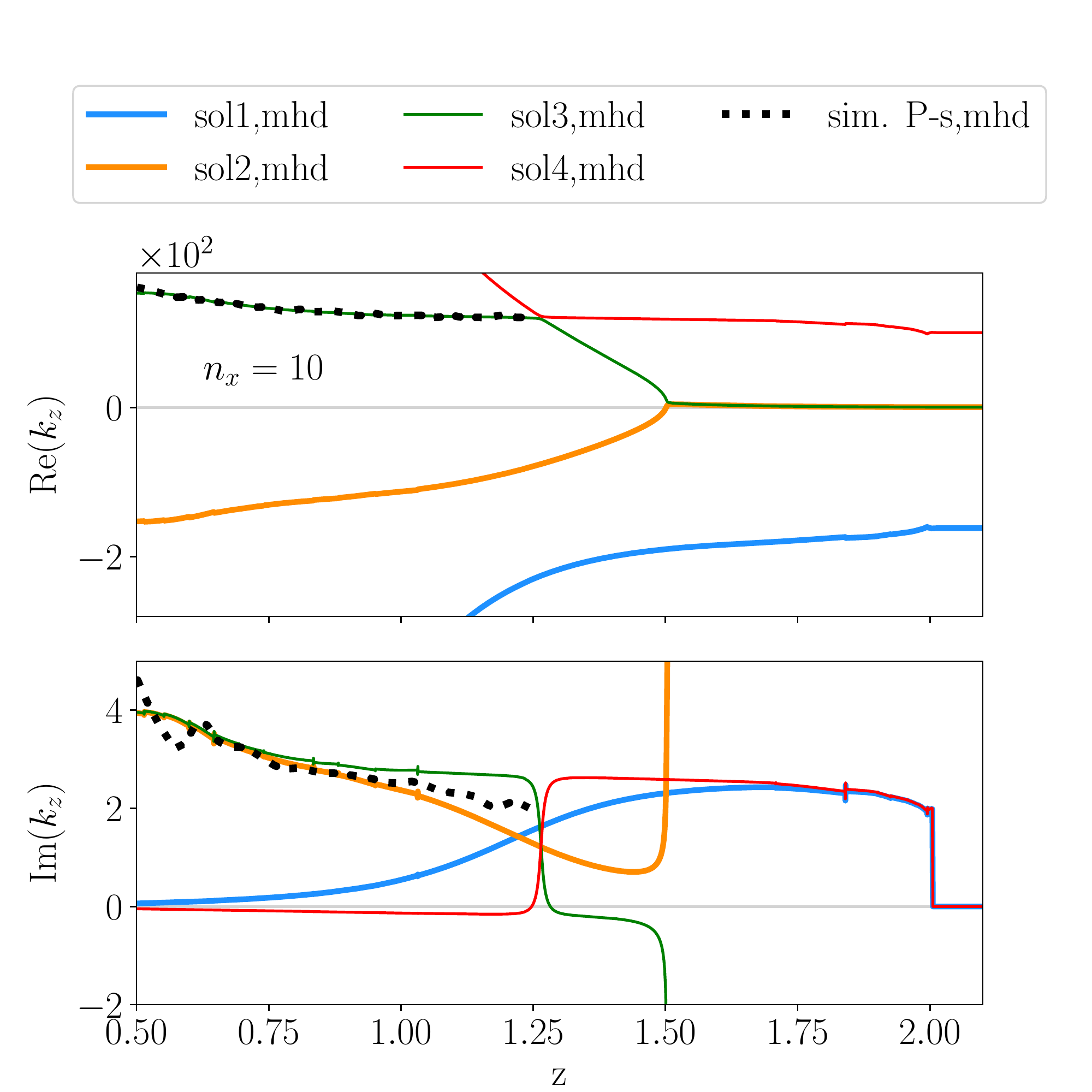}}
\caption{
Left: Profile of vertical velocity along $x=0$ for P-s.  Right: Local dispersion relation corresponding to the simulations shown in Figure \ref{fig:snap_g2}.
The complex vertical wavenumber $k_z(z)$ values calculated from the P-s simulation are overplotted on the dispersion diagram with the black dotted lines, up to the equipartition height, and they are clearly in perfect agreement with the upward fast wave (in green). 
}
\label{fig:psd}
\end{figure*}
%%%%%%%%%%%%%%%%%%%%%%%%%%%%%%%%%%%%%%%%%
\begin{figure*}[htb]
\centering
\FIG{\includegraphics[width=0.8\textwidth]{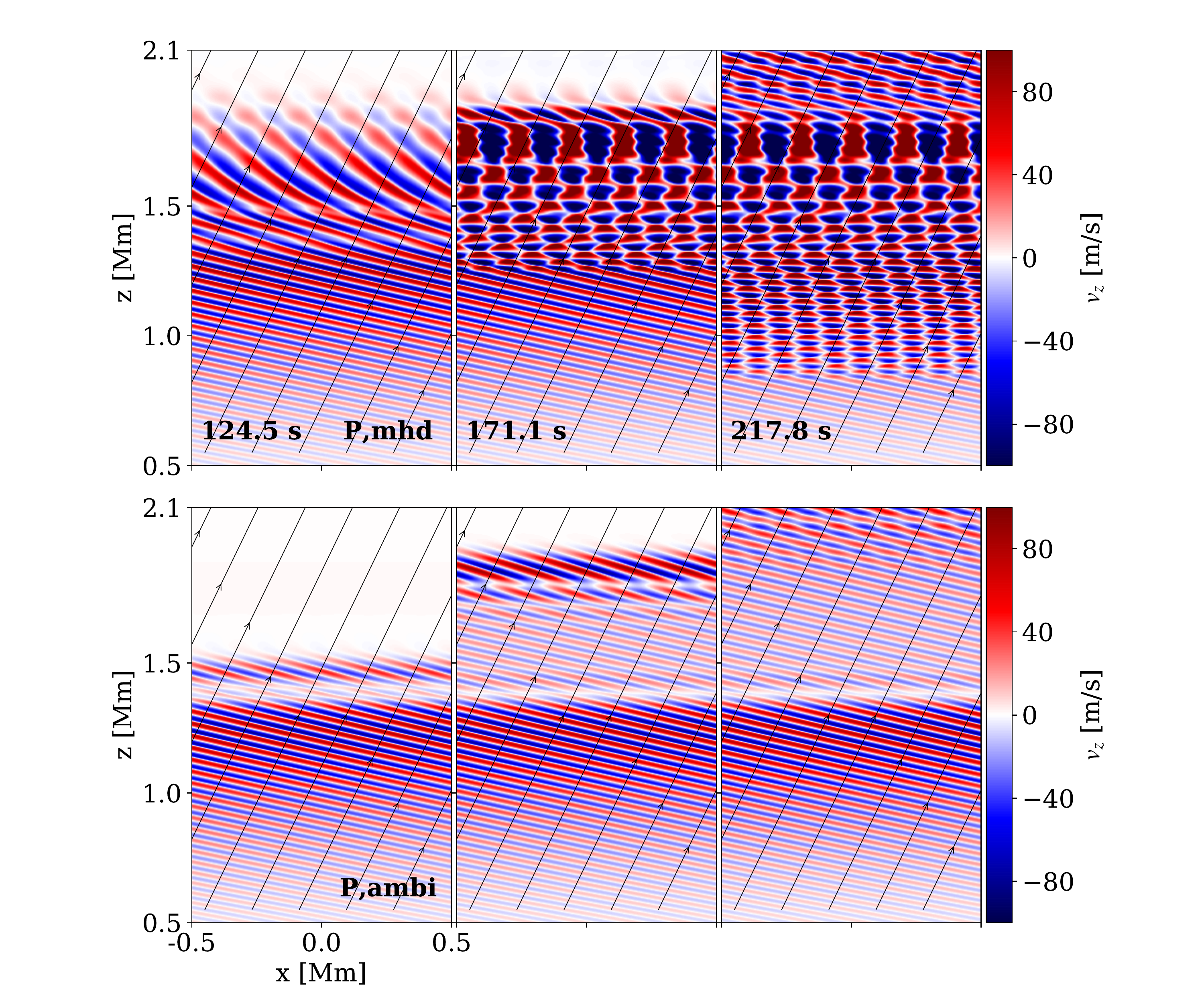}}
\caption{
Snapshots of vertical velocity for plane wave case P in the MHD case (top row) and ambipolar case (bottom row) at three different moments. Note the difference in interference patterns, almost absent for the ambipolar case.
}
\label{fig:snapP}
\end{figure*}
%%%%%%%%%%%%%%%%%%%%%%%%%%%%%%%%%%%%%%%%%
\begin{figure*}[htb]
\centering
\FIG{\includegraphics[width=0.8\textwidth]{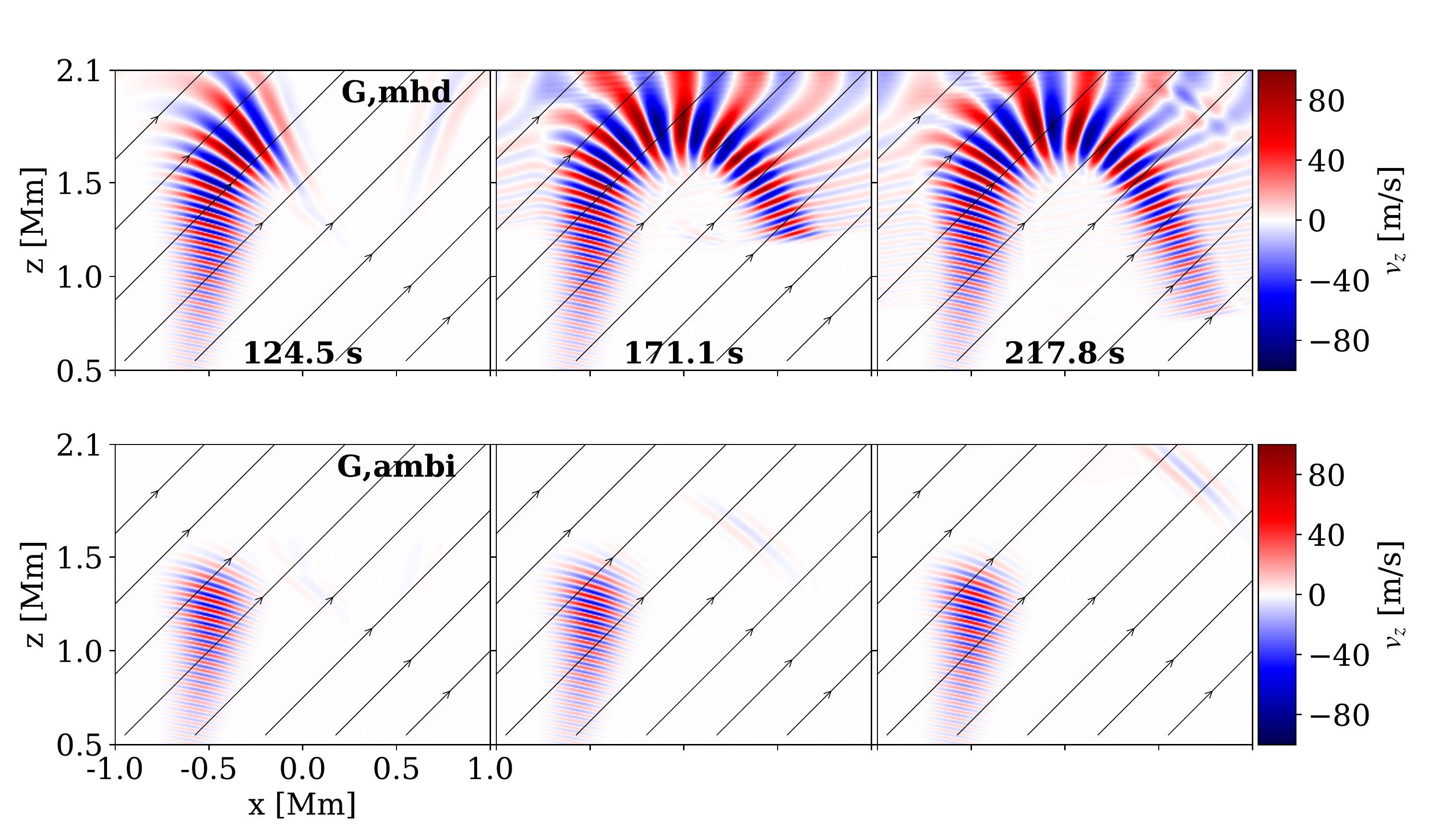}}
\caption{
Snapshots of vertical velocity for the gaussian case G simulation in the MHD case (top row) and ambipolar case (bottom row) at three different moments. The reflected wave is almost absent in the ambipolar run.
}
\label{fig:snapG}
\end{figure*}

%%%%%%%%%%%%%%%%%%%%%%%%%%%%%%%%%%%%%%%%%
\begin{figure*}[htb]
\centering
\FIG{\includegraphics[width=0.8\textwidth]{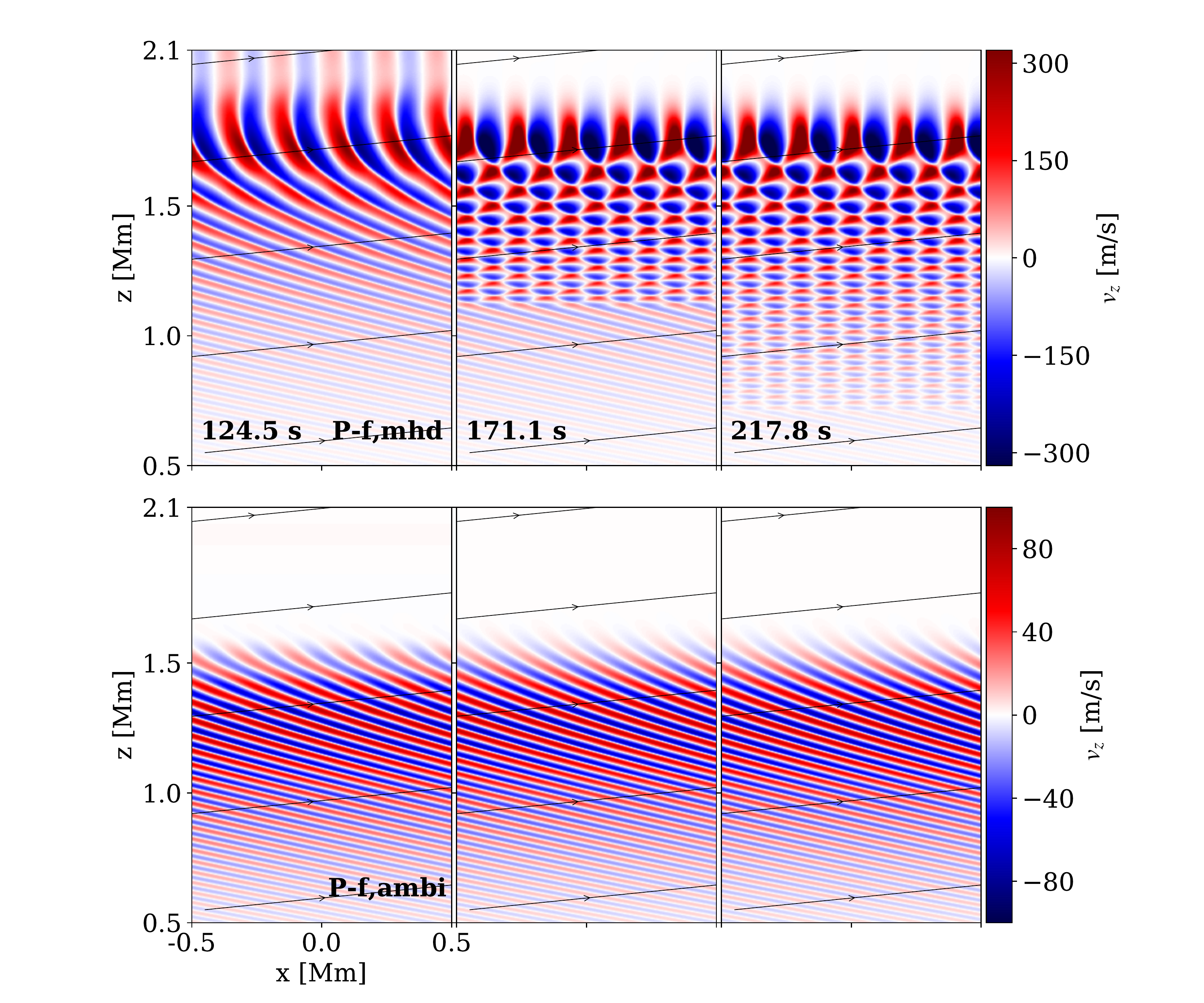}}
\caption{
Snapshots of vertical velocity for P-f simulation in the MHD case (top row) and ambipolar case (bottom row) at three different moments. Note the difference in the interference patterns, almost absent for the ambipolar case.
}
\label{fig:snapPf}
\end{figure*}
%%%%%%%%%%%%%%%%%%%%%%%%%%%%%%%%%%%%%%%%%
\begin{figure*}[h]
\centering
\FIG{\includegraphics[width=8cm]{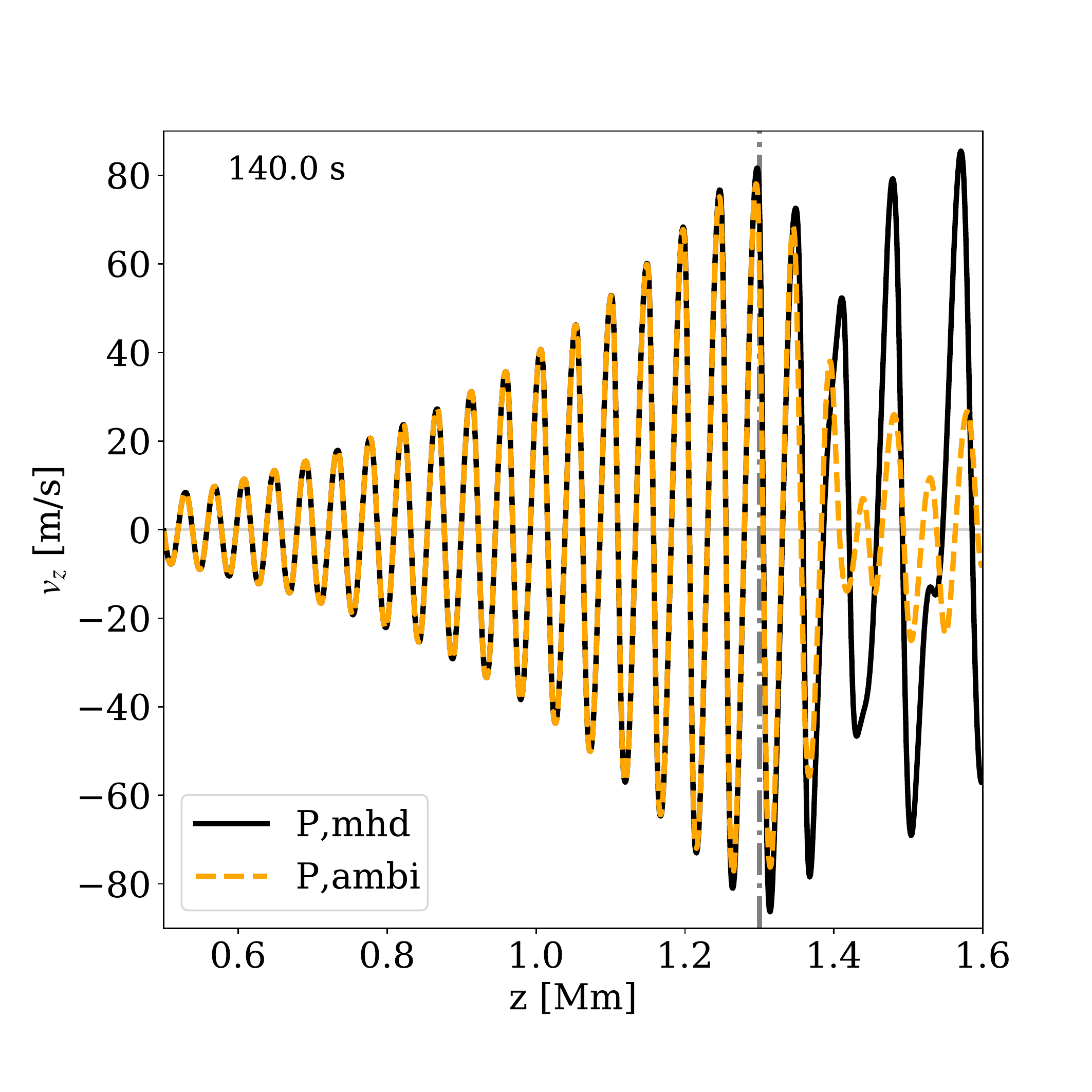}
\includegraphics[width=8cm]{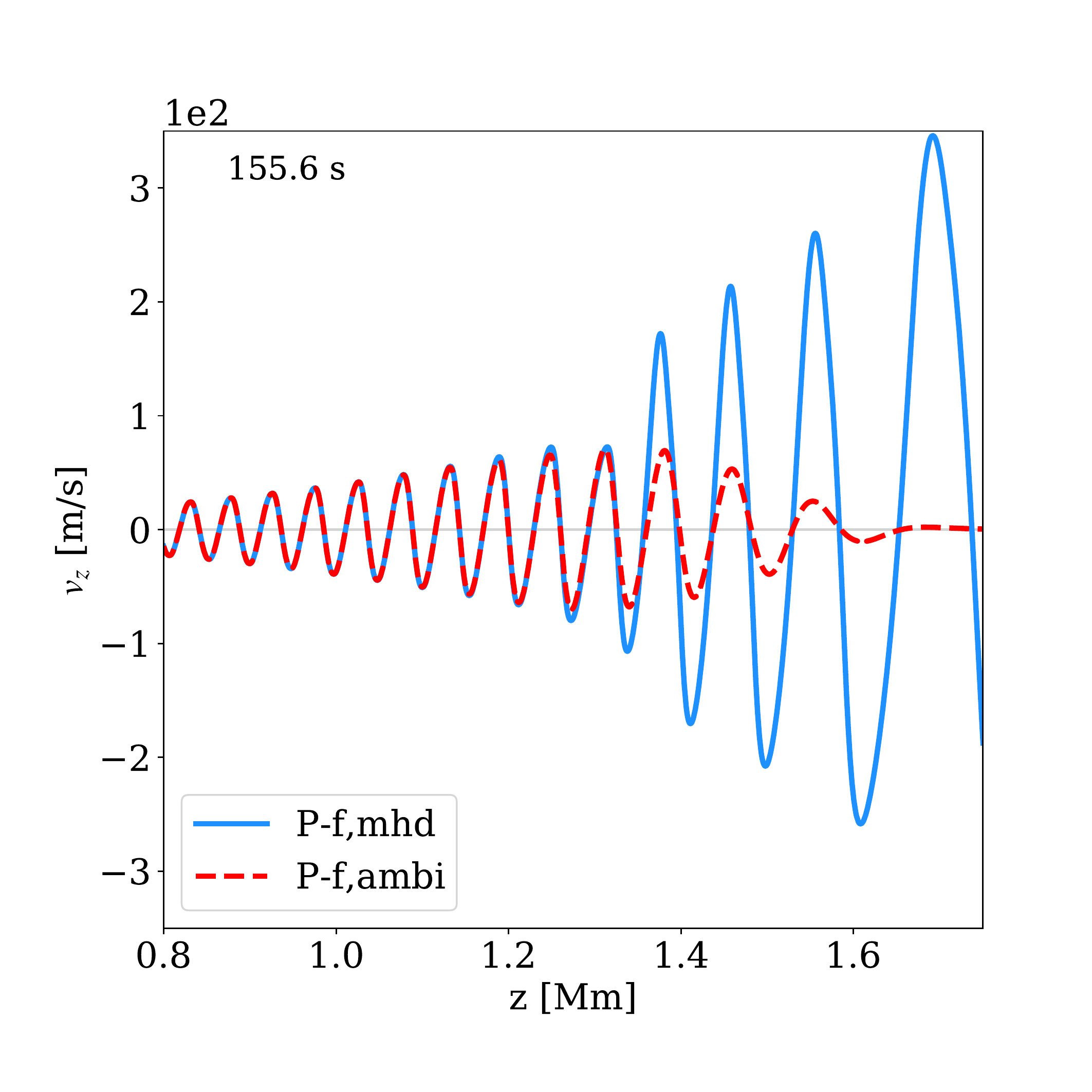}}
\caption{
Vertical cuts at $x=0$ for the vertical velocity, for both the MHD and the ambipolar runs from Figs.~\ref{fig:snapP}-\ref{fig:snapPf}. Left: the plane wave case P. 
{ The vertical dot-dashed gray line located at the height $z=1.3$ Mm marks the equipartition layer.}
Right: the fast plane wave case P-f.
}
\label{fig:p1}
\end{figure*}
%%%%%%%%%%%%%%%%%%%%%%%%%%%%%%%%%%%%%%%%%
%%%%%%%%%%%%%%%%%%%%%%%%%%%%%%%%%%%%%%%%%%%%%%%%
\begin{figure*}[h]
\FIG{\includegraphics[width=8cm]{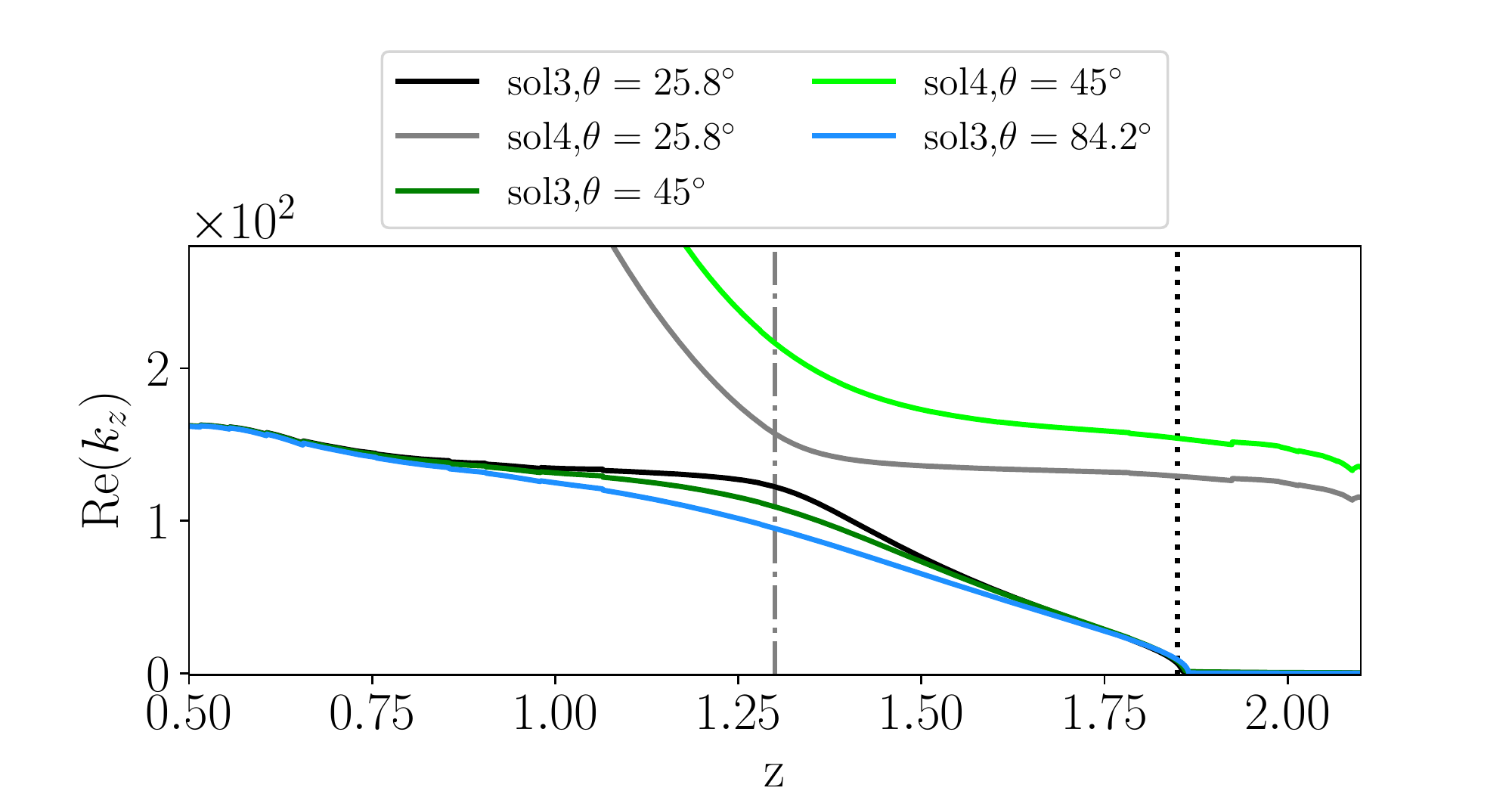}
\includegraphics[width=8cm]{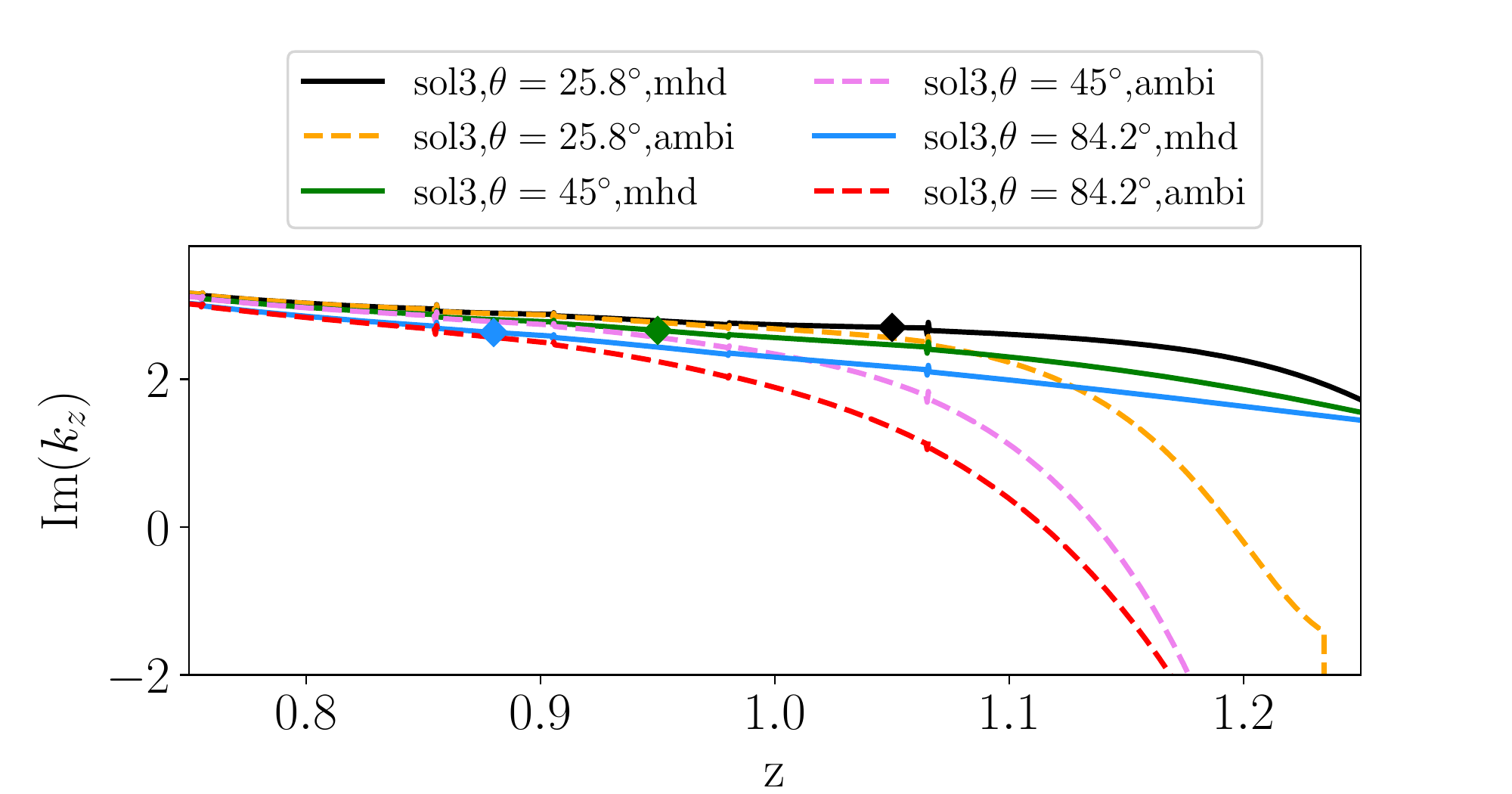}}
\caption{
Left: Solutions 3 and 4 for the real part of the vertical wavenumber $k_z$ corresponding to the cases P ($\theta=25.8^\circ$),
and G ($\theta=45^\circ$), as well as solution 3 for case P-s.
The vertical dot-dashed gray line at $z=1.3$ Mm, also present in left panel of Figure \ref{fig:p1} marks the equipartition layer.
The vertical dotted black line at $z=1.85$ Mm shows the reflection height.
Right: Solution 3  for the imaginary part of the vertical wavenumber $k_z$ corresponding to the three cases P ($\theta=25.8^\circ$),
G ($\theta=45^\circ$) and P-f ($\theta=84.2^\circ$), for both MHD and ambipolar cases. 
The points where the imaginary part  diverges in the ambipolar case compared to the MHD case are indicated with markers in the panel.
}
\label{fig:p2}
\end{figure*}
%%%%%%%%%%%%%%%%%%%%%%%%%%%%%%%%%%%%%%%%%%%%%%%%%%%%%%%%%
%%%%%%%%%%%%%%%%%%%%%%%%%%%%%%%%%%%%%%%%%%%
\begin{figure*}[!htb]
\centering
\FIG{
\includegraphics[width=8cm]{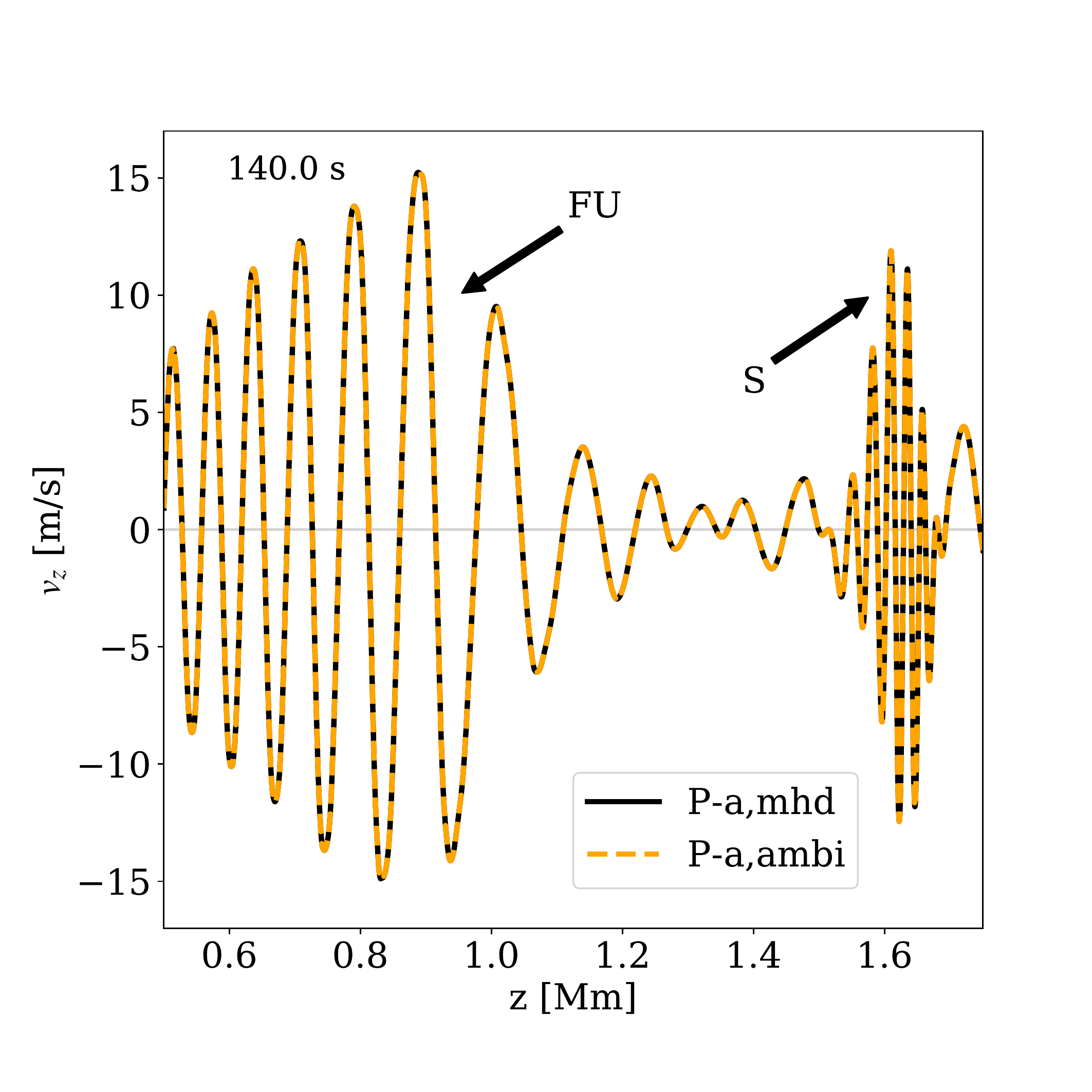}
\includegraphics[width=8cm]{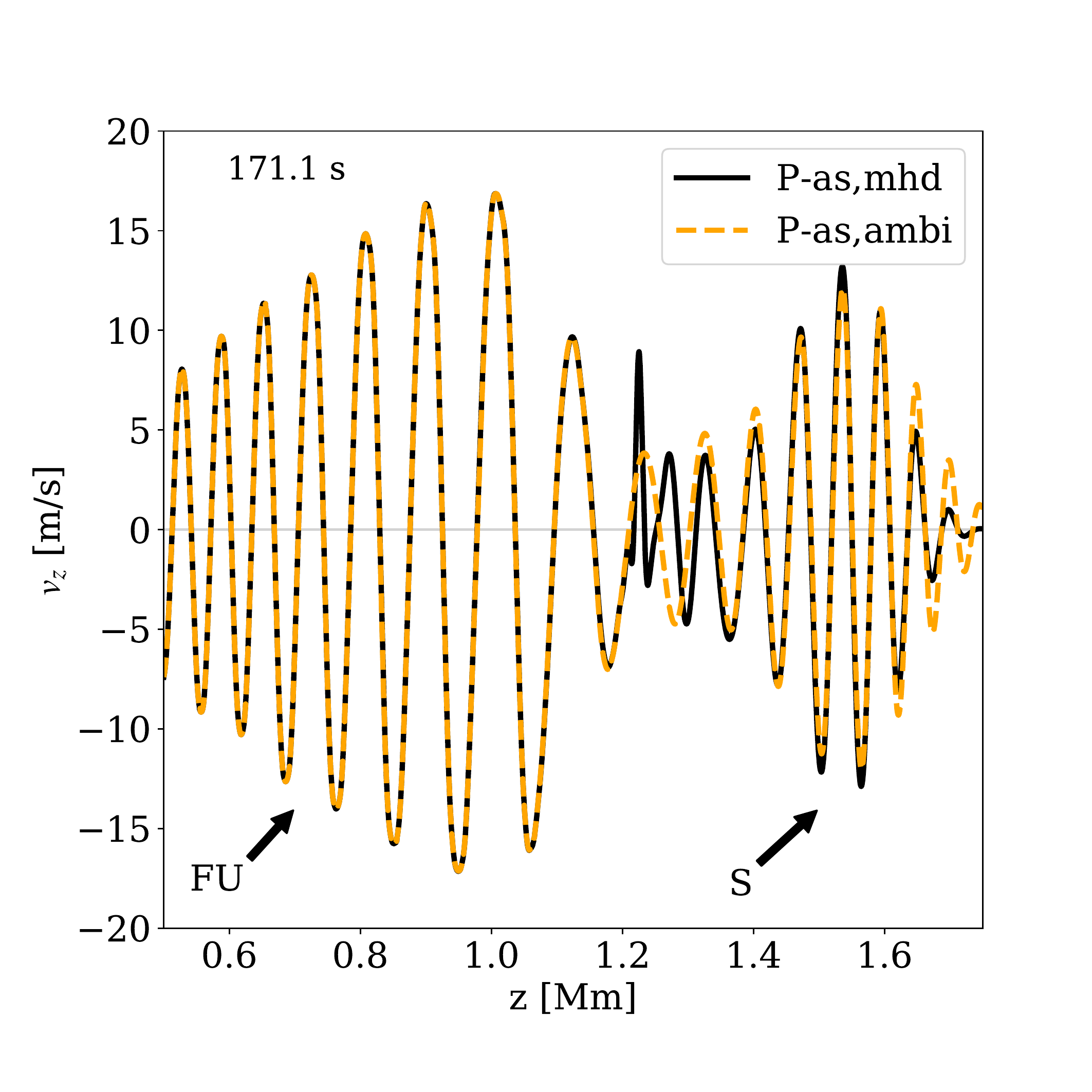}
}
\caption{
Vertical cut at $x=0$ for the vertical velocity for plane wave case P-a (left panel)
and P-as (right panel). The upward fast (FU) and the slow (S) components are indicated in both panels.
We compare MHD (black solid lines)  with the ambipolar (orange dashed lines)  case.
}
\label{fig:Pa}
\end{figure*}
%%%%%%%%%%%%%%%%%%%%%%%%%%%%%%PERIOD
%%%%%%%%%%%%%%%%%%%%%%%%%%%%%%%%%%%%%%%%%%%
\begin{figure*}[!htb]
\centering
\FIG{\includegraphics[width=8cm]{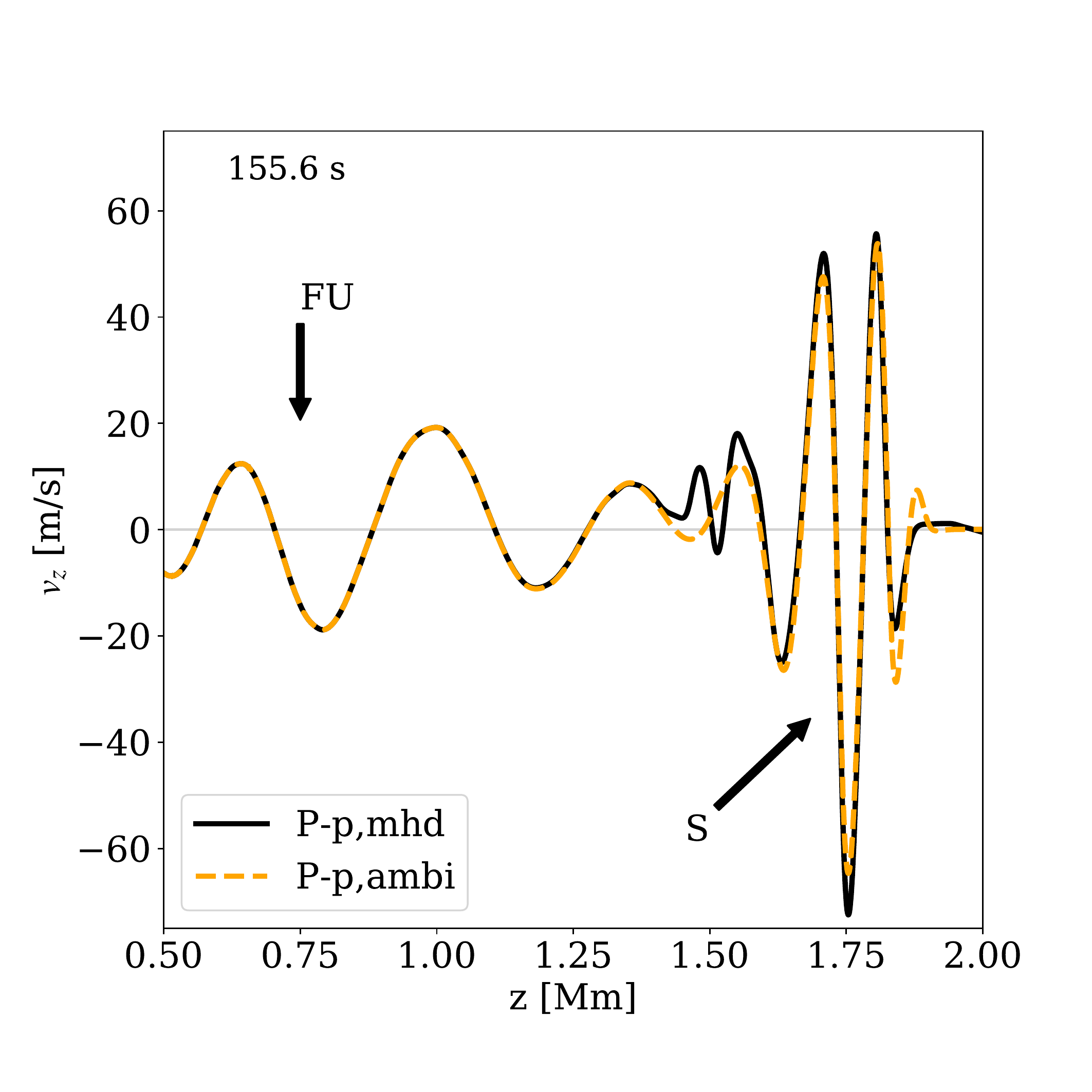}
\includegraphics[width=8cm]{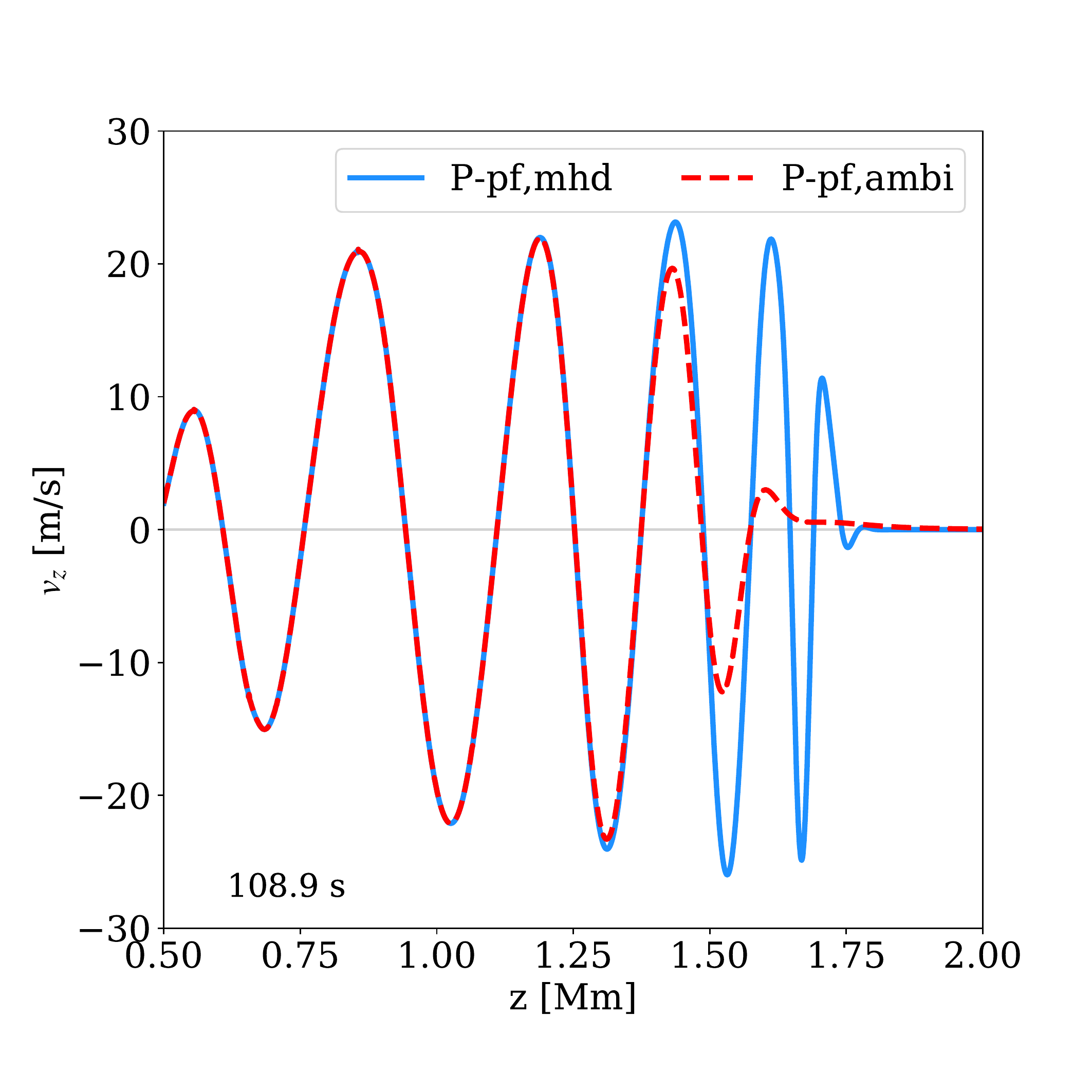}}
\caption{
Vertical cut at $x=0$ for the vertical velocity, for both MHD and ambipolar cases. Left: case P-p, the upward fast (FU) and the slow (S) components are indicated in 
the left panel.
Right: P-pf.
}
\label{fig:Pp}
\end{figure*}
%%%%%%%%%%%%%%%%%%%%%%%%%%%%%%%%%%%%%%%%%%%%%%%%%%%%%%%%%%% 
%%%%%%%%%%%%%%%%%%%%%%%%%%%%%%%%%%%%%%%%%%%

\begin{figure*}[!htb]
\centering
\FIG{\includegraphics[width=8cm]{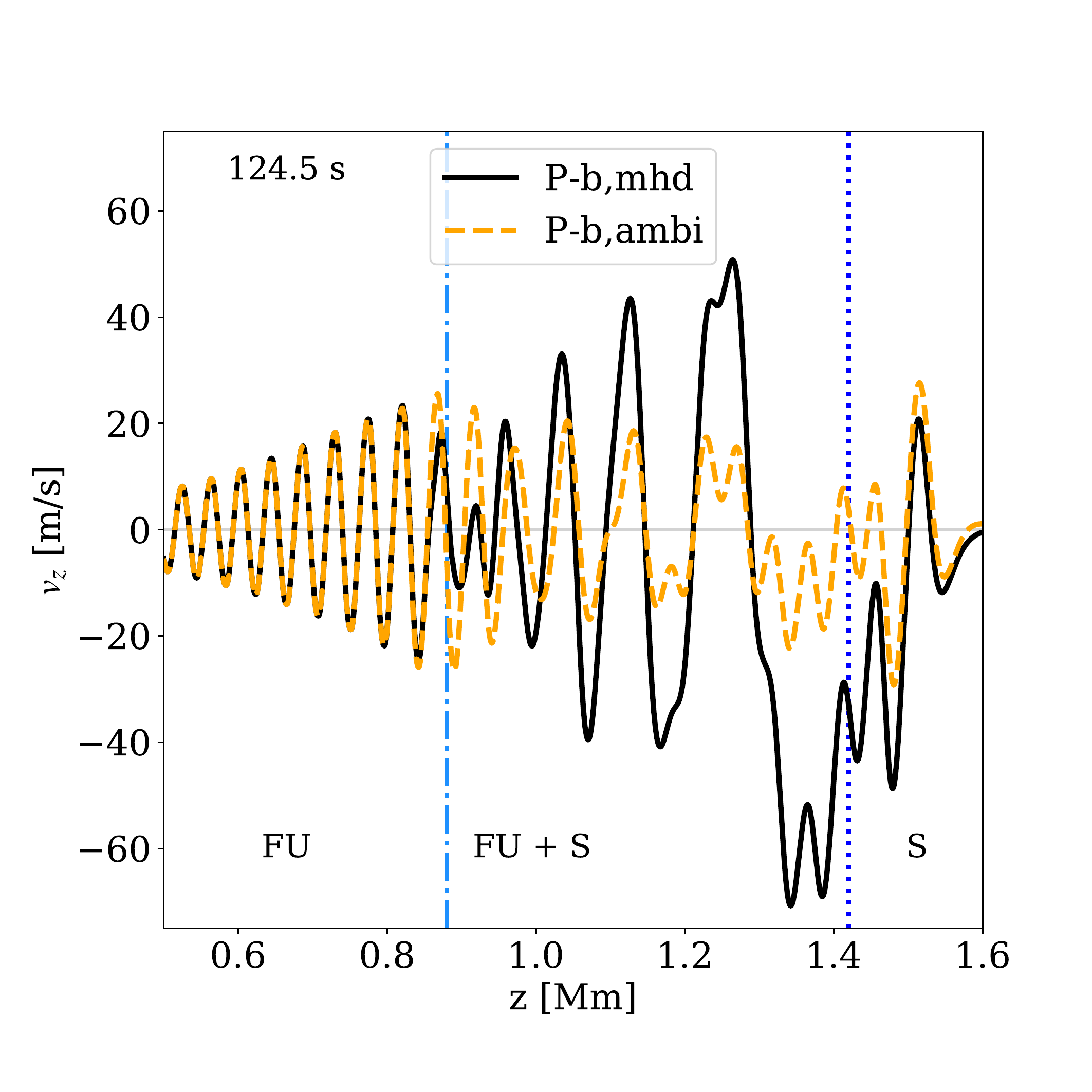}
\includegraphics[width=8cm]{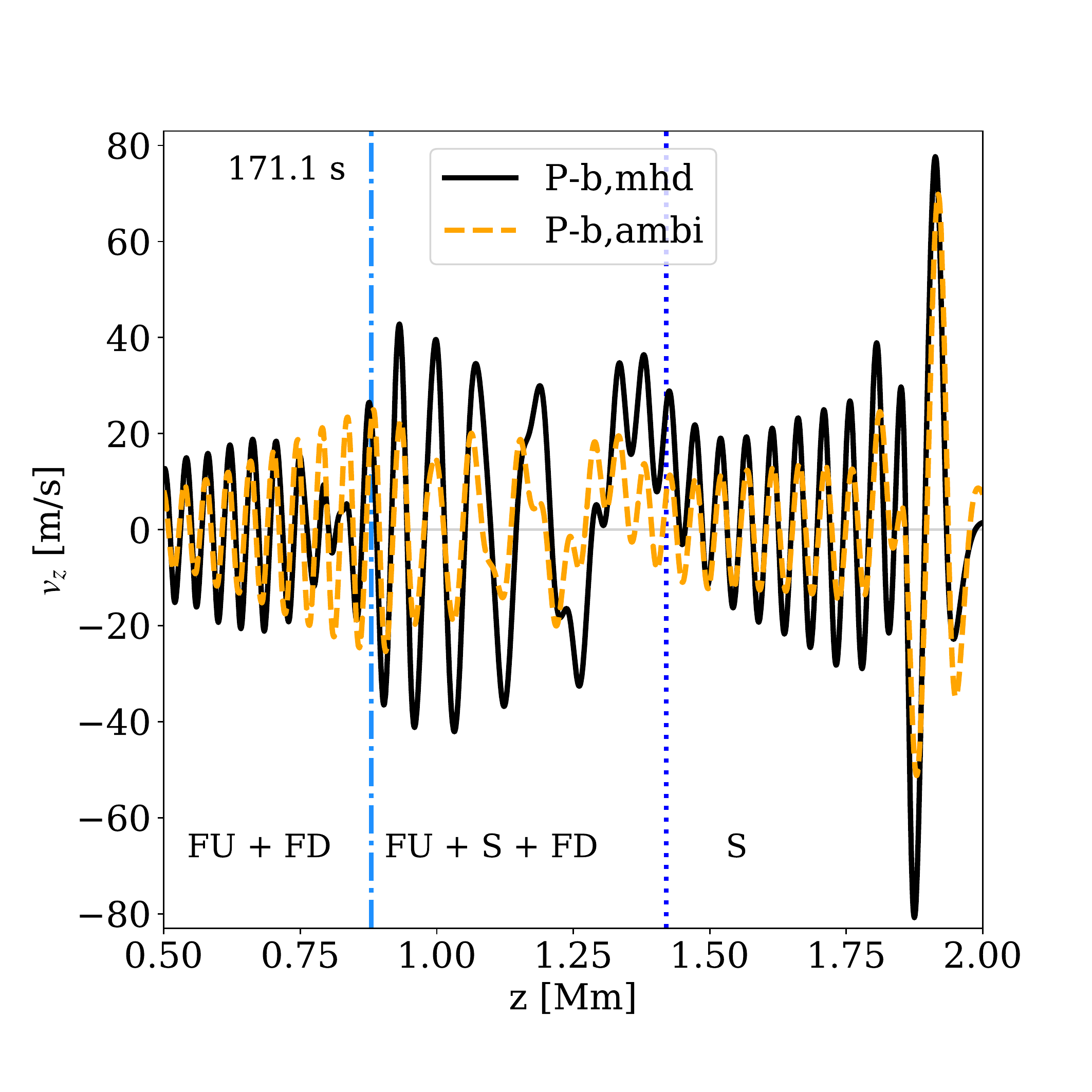}}
\caption{
Vertical cuts at $x=0$ for the vertical velocity for the simulation P-b at two different times (note the changing axis scales). 
This can be compared with the weaker field strength case from Fig.~\ref{fig:p1}, left panel. 
The equipartition height is shown by a vertical light blue dotted line located at $z=0.85$ Mm in both panels. 
In the right panel the vertical dotted dark blue line located at $z=1.42$ Mm
indicates the reflection height of the fast component. The wave modes: upward fast (FU), slow (S) and downward fast (FD) present in each of the regions are indicated in the panels.
}
\label{fig:Pb}
\end{figure*}
%%%%%%%%%%%%%%%%%%%%%%%%%%%%%%%%%%%%%%%%%%%
\begin{figure*}[!htb]
\centering
\FIG{\includegraphics[width=8cm]{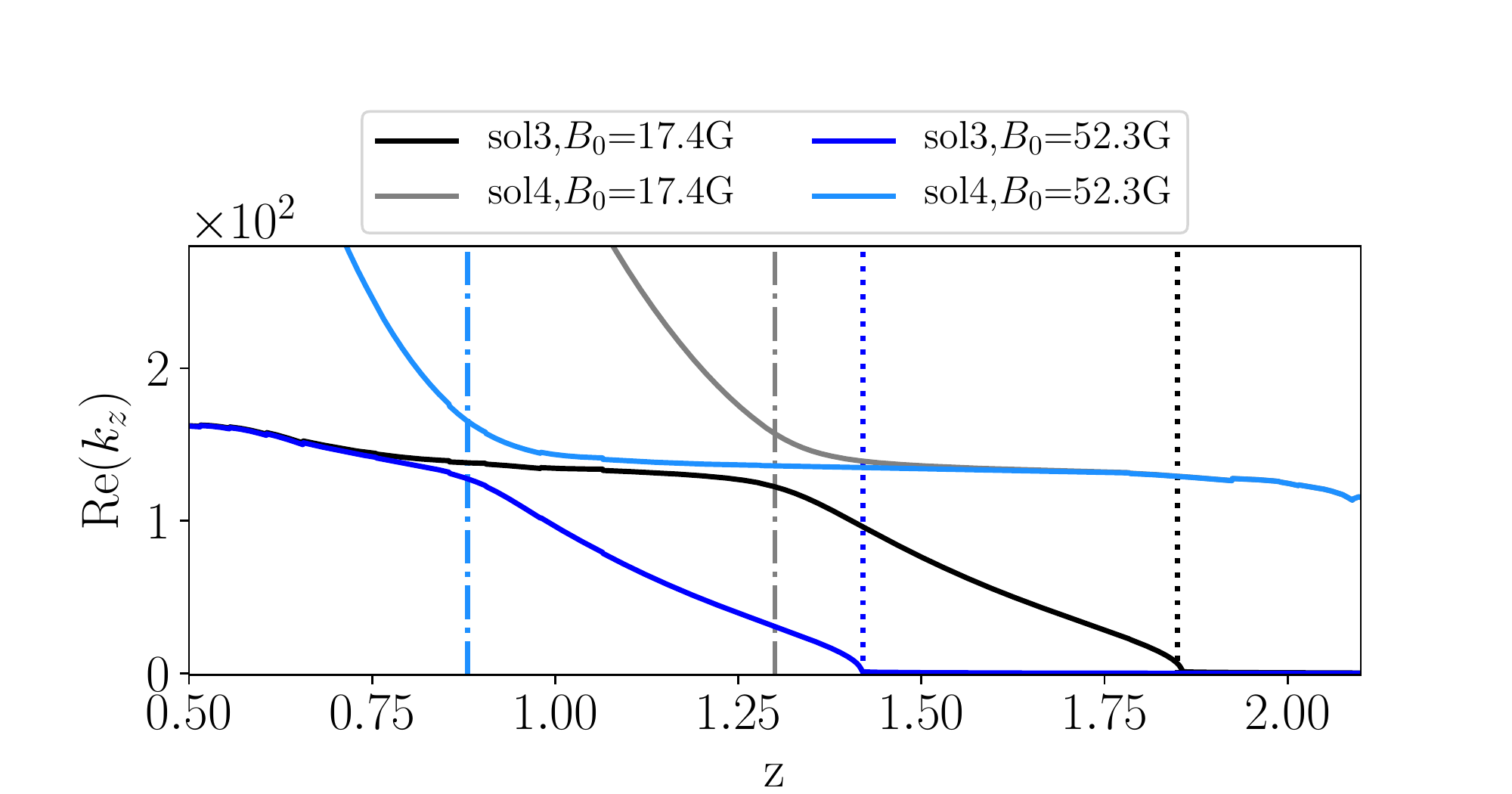}
\includegraphics[width=8cm]{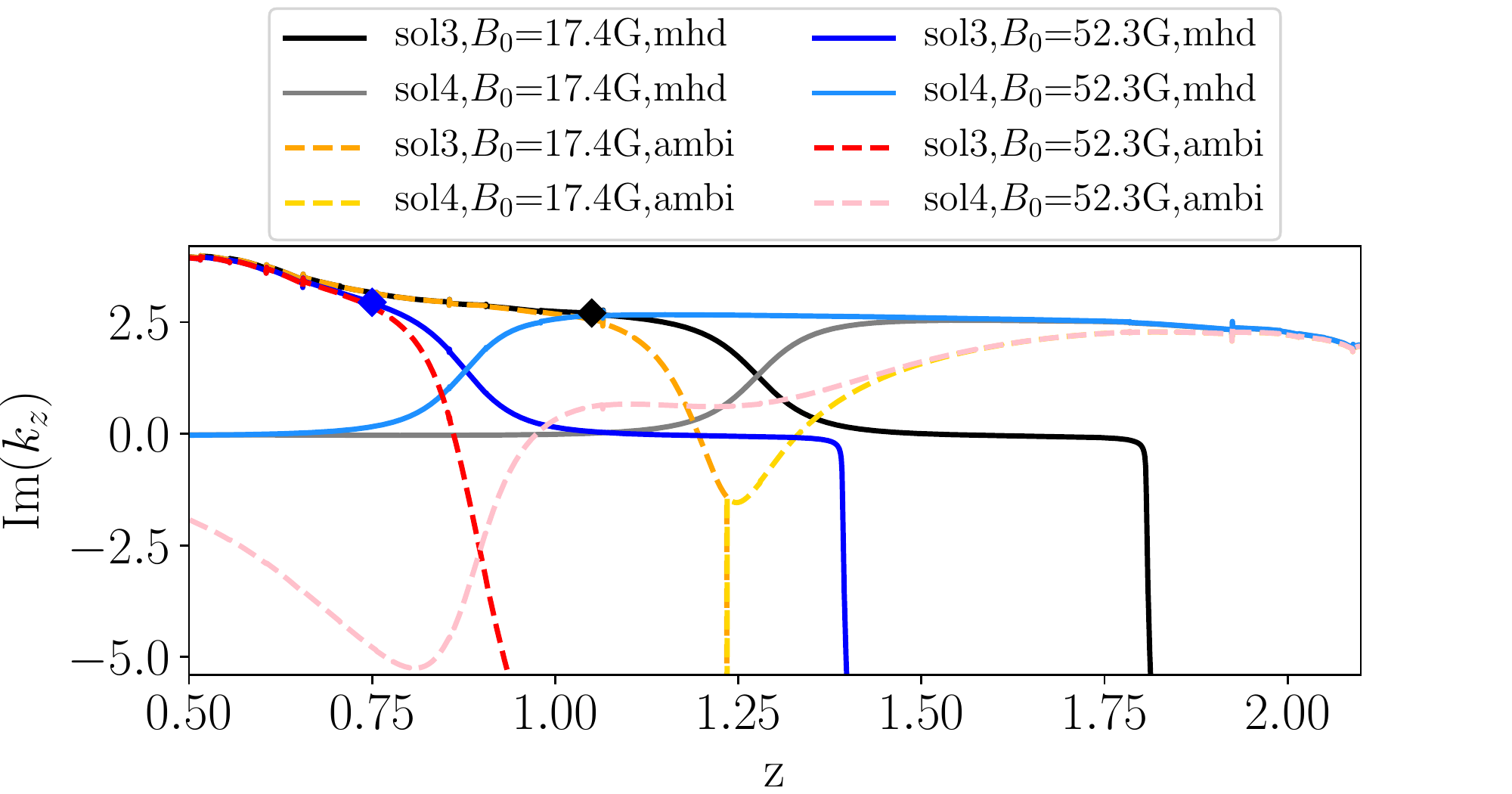}}
\caption{
Relevant solutions 3 and 4 of the local dispersion relation for the real (left) and imaginary (right) part of the vertical wavenumber $k_z$ corresponding to the cases P ($B_0=17.4G$) and
P-b ($B_0=52.3$). The MHD solutions are indicated by solid lines in both panels. 
In the left panel the reflection heights for the fast component are indicated by  black and blue vertical dotted lines for P and P-b, respectively.
The equipartition heights are shown with a gray and light blue vertical dot-dashed lines for P and P-b, respectively.
In the right panel we also show the solutions 3 and 4 for the ambipolar cases with dashed lines.
The points where the ambipolar and MHD solution 3 (fast modes) diverge are indicated by the markers for both P (black marker) and P-b (blue marker) cases.  
}
\label{fig:dispB}
\end{figure*}
%%%%%%%%%%%%%%%%%%%%%%%%%%%%%%%%%%%%%%%%%%%

In order to have an analytical understanding of the simulations performed in this study we will use the dispersion relations~(\ref{eq:disp})
with the coefficients corresponding to the MHD and ambipolar situation. 
%The first order corrections, when the gradients in the amplitude and the vertical wave number
%are considered first order terms, as described in \citep{Popescu2019}, were several orders of magnitude smaller and are not used in this study.
In our Appendix~\ref{appendixA}, we compare our local dispersion relation approach to the WKB analysis obtained by \cite{Paul1}
and to another approximate dispersion relation suggested by~\cite{thomas1}.  We find very good agreement between the real parts of the complex wavenumbers $k_z$, which determine the phase speeds of our fast and slow, upward and downward propagating waves. The imaginary parts, which relate to the damping of the waves as they propagate, are found to differ, but we will compare our nonlinear simulation results with our local predictions.
%%%%%%%%%%%%%%%%%%%%%%%%%%%%%%%%%%%%%%%%%%%%%%%%%%%%%%%%%%%%%%%%%%%%%%%%%%%
%%%%%%%%%%%%%%%%%%%%%%%%%%%%%%%%%%%%%%%%%%%%%%%%%%%%%%%%%%%%%%%%%%%%%%%%%%%
\section{2D simulation results}   \label{sec:res}
%%%%%%%%%%%%%%%%%%%%%%%%%%%%%%%%%%%%%%%%%%%%%%%%%%%%%%%%%%%%%%%%%%%%%%%%%%%
%
%%%%%%%%%%%%%%%%%%%%%%%%%%%%%%%%%%%%%%%%%%%%%%%%%%%%%%%%%%%%%%%%%%%%%%%%%%%
\subsection{Ideal MHD preliminaries}
%%%%%%%%%%%%%%%%%%%%%%%%%%%%%%%%%%%%%%%%%%%%%%%%%%%%%%%%%%%%%%%%%%%%%%%%%%%
%
%%%%%%%%%%%%%%%%%%%%%%%%%%%%%%%%%%%%%%%%%%%%%%%%%%%%%%%%%%%%%%%%%%%%%%%%%%%
\subsubsection{Plane waves versus gaussian pulse}
%%%%%%%%%%%%%%%%%%%%%%%%%%%%%%%%%%%%%%%%%%%%%%%%%%%%%%%%%%%%%%%%%%%%%%%%%%%
%
Figure \ref{fig:snap_g2} shows snapshots of the vertical velocity for the simulations P-s, G-s, G-sr from Table~\ref{tab:sims}.
The waves driven for P-s and G-s have the same parameters, the only difference being in the type of perturbation, being a plane wave for P-s
and a gaussian for G-s. 
The horizontal mode number used for P-s and G-s is $n_x=10$, the horizontal wave number is calculated as $k_x = 2 \pi n_x/L_x$,
with the horizontal box length $L_x$ given in Table \ref{tab:sims}, making the angle $\phi = 22.3^\circ$. 
For the  G-sr wave we have $n_x=-10$, reversing the orientation of the wavevector. 

When  the upward traveling fast  waves reach the equipartition layer,  they  usually split into two components, consisting of a fast wave which will be reflected
and a slow mode which is transmitted higher up in the atmosphere. 
{
The transmission coefficient is shown to be \citep{Paul1new}
\begin{equation}\label{eq:tra}
T \propto \text{exp}\left( -\pi k h_s \sin^2(\alpha) \right)\,,
\end{equation}
where $h_s$ is equipartition layer scale height, which is the thickness of the layer over which $v_{\rm A0} \approx c_0$, measured along the direction of propagation $\mathbf{k}$. 
}
In Eq.~(\ref{eq:tra}), $\alpha$ is the angle between the direction of propagation
of the wave and the magnetic field, so in our notation it is $\alpha=\mid \theta - \phi\mid$. 
The angle $\alpha$ for G-s and P-s corresponds to a wave which propagates almost parallel to the magnetic field.
For this reason, the transmission from the fast to a slow wave is almost complete. 
This can be better seen in the snapshots for the gaussian perturbation, G-s, where only a small fraction of the upward wave
seems to be reflected at the height around 1.5 Mm. Note that the gaussian wave is really a superposition of plane waves with wave numbers centered at the wave numbers corresponding
to $n_x=10$, and different components reflect at slightly different heights, as can be observed in the snapshots for G-s (middle row of Fig.~\ref{fig:snap_g2}).
Because of the larger corresponding angle $\alpha$, the reversed gaussian case G-sr is reflected. 
We can observe in Figure \ref{fig:snap_g2} that the propagation speed and the reflection height of G-sr are similar to that of the reflected part of the G-s wave (bottom and middle rows).

Figure \ref{fig:gs} shows the profile of the vertical velocity along the green line shown in the right panels of Figure \ref{fig:snap_g2}.
We can observe that the  plane wave case P-s and the gaussian case G-s have similar profiles, with the only difference being a smaller amplitude for G-s because of the spreading suffered by the gaussian pulse.
The profile of the vertical velocity for the reversed gaussian G-sr case is also similar to the profile of P-s and G-s in the bottom part of the atmosphere. 
This is because the bottom part corresponds to a weak field regime
($v_{A0} \ll c_0$) and the fast waves are then acoustic in nature, so they propagate with approximately the sound speed, regardless of the propagation angle and the magnetic field orientation.
Higher up in the atmosphere, the propagation speed for G-sr becomes  larger than for G-s (and P-s), and the amplitude of the vertical velocity drops, becoming negligible at the height $z\approx 1.5$ Mm.
%
%%%%%%%%%%%%%%%%%%%%%%%%%%%%%%%%%%%%%%%%%%%%%%%%%%%%%%%%%%%%%%%%%%%%%%%%%%%
\subsubsection{Comparing the simulation with the local dispersion relation}
%%%%%%%%%%%%%%%%%%%%%%%%%%%%%%%%%%%%%%%%%%%%%%%%%%%%%%%%%%%%%%%%%%%%%%%%%%%
%
Left panel of Figure \ref{fig:psd} shows the vertical velocity along the vertical line $x=0$ for the plane wave case P-s. We use this vertical profile in order to obtain the real and imaginary part of the local `vertical wave number' $k_z(z)$ corresponding
to this wave. We use the positions of the maximum and minimum peaks, from which we determine the local wavelength and amplitude variation. Afterwards, we interpolate 
in order to obtain values in the whole vertical domain. 

The right panel of Figure \ref{fig:psd} shows the four solutions of the ideal MHD, local dispersion relation from Eq.~(\ref{eq:disp}) for this case.
We overplotted the real and imaginary part obtained from the simulation P-s up to the equipartition height.
We observe that there is a good match for both real and imaginary part between the simulation and the solution corresponding to the fast upward wave of our local dispersion relation (in Figure \ref{fig:psd}, the green line for solution 3).
At the equipartition layer (at $\approx$ 1.25 Mm), there is a crossing of two solution branches, namely between the upward transmitted slow wave (red) and the fast wave (green). The fast wave is reflected
above this point (at $\approx$ 1.5 Mm, we see the green line in Figure \ref{fig:psd} showing an infinite phase speed $\omega/\text{Re}(k_z)$ and a pure damping).  
%We observe that the positive peak on the net velocity in the left panel is located at the reflection point.
We observe that the reflection height at $z=1.5$ Mm is consistent with its visual estimation from Figure \ref{fig:snap_g2}.

The upward solutions of the dispersion relation for the reversed gaussian G-sr ($n_x=-10$) could be also deduced from the dispersion diagram in the right panel of Figure \ref{fig:psd} for the $n_x=10$ case. Indeed, the upward slow and the fast wave corresponding to $n_x=-10$ have for the real part a symmetric image in the positive domain 
and the same imaginary part as for the solutions 1 (in blue) and 2 (in orange) for $n_x=10$, respectively. 
In the dispersion diagram, we can visually see the transmission coefficient as the distance between the two solutions describing the fast and the slow waves at the equipartition layer.
The fact that P-s and G-s are mostly transmitted compared to G-sr then appears as the fact that curves that represent the real part of the solutions 3 and 4 join at $z\approx 1.25$ Mm, compared
to the larger distance seen between the curves at that position corresponding to the solutions 1 and 2.
Thus, in order to compare G-sr to G-s (and P-s) from this diagram, we have to compare the orange line (by looking at the absolute value for the real part of $k_z$)
to the green line up to the equipartition height ($z\approx1.25$ Mm) and to the red curve above it.  
The larger propagation speed and smaller amplitude for G-sr compared to G-s (and P-s) above $z\approx1$ Mm can be deduced from this diagram.
%%%%%%%%%%%%%%%%%%%%%%%%%%%%%%%%%%%%%%%%%%%%%%%%%%%%%%%%%%%%%%%%%%%%%%%%%%%
\subsection{Ideal MHD versus ambipolar cases}
%%%%%%%%%%%%%%%%%%%%%%%%%%%%%%%%%%%%%%%%%%%%%%%%%%%%%%%%%%%%%%%%%%%%%%%%%%%
%%%%%%%%%%%%%%%%%%%%%%%%%%%%%%%%%%%%%%%%%%%%%%%%%%%%%%%
%
\subsubsection{Varying the magnetic field inclination}
{ Compared to the simulation P-s presented above, }
we next change the angle of propagation for the waves to a smaller inclination, $\phi=10.9^\circ$,  
and then we change progressively the angle of inclination of the magnetic field with the vertical direction from $\theta=25.8^\circ$ (plane wave P simulation),
$\theta=45^\circ$ (gaussian pulse G) and $\theta=84.2^\circ$ (plane wave case P-f). In this section, we will compare ideal MHD with cases where
we also take into account the ambipolar effects.

Figures \ref{fig:snapP}, \ref{fig:snapG} and \ref{fig:snapPf} show snapshots of P, G, and P-f simulations at different moments of time. Top and bottom panels show the MHD and ambipolar cases, respectively. The wave in the G case is mostly reflected.
We can observe in all three cases that the wave amplitude is damped when the ambipolar diffusion is taken into account. For the G case, the ambipolar run shows therefore hardly any reflected wave. The interference patterns caused by interacting up and downward wave branches are clearly different between ideal MHD and ambipolar runs, for P and P-f simulations.
The effect of having a smaller inclination of the wave propagation with the vertical direction
is an increase of the reflection height, which is now located at $z\approx1.8$ Mm.

Left and right panels of Figure \ref{fig:p1} show the profiles $v_z(z)$ of the vertical velocity at the vertical cut $x=0$ for the plane wave P and P-f cases, respectively.
The snapshots considered are taken before the  waves reach the reflection point, so that we can easily distinguish between the upward and downward waves.
In the left panel the equipartition layer is marked by a vertical gray dotted line located at $z=1.3$ Mm.
We can observe interference for the P case  above the equipartition layer, which at this moment can only be created 
by the fast and the slow waves propagating upwards.
The transmission to the slow wave is decreased in the P case, because  $\alpha$ from the above Eq. (\ref{eq:tra}) increased,
compared to the simulation P-s presented above. Therefore, the amplitude of the fast upward
wave which will be reflected above the equipartition layer is larger than for the previous P-s case and the interference between the slow and the fast waves is clearly visible. 
In contrast, there is no visible interference for the case P-f, and the wave seems to go completely on the fast branch. The interference patterns seen for P and P-f at later stages in Figures \ref{fig:snapP} and \ref{fig:snapPf} are related to the interaction between the upward 
fast wave driven from below, the downward reflected fast wave, and additionally the upward slow wave transmitted at the equipartition layer for the P case.
The damping due to the ambipolar diffusion of the fast upward mode appears at earlier height and is larger for P-f compared to P case. 
%
%%%%%%%%%%%%%%%%%%%%%%%%%%%%%%%%%%%%%%%%%%%%%%%%%%%%%%%
\subsubsection{Analysis using the local dispersion relation}
%%%%%%%%%%%%%%%%%%%%%%%%%%%%%%%%%%%%%%%%%%%%%%%%%%%%%%%
%
We observe from the left panel of Figure \ref{fig:p2}, which shows the real part of the relevant dispersion relation solutions, that the three waves (P, G and P-f) reflect at the same point, 
namely at $\approx$ 1.85 Mm where $\text{Re}(k_z)$ vanishes, which is higher up in the atmosphere than in the previous P-s case.
The reflection height, { marked by a vertical black dotted line at $z=1.85$ Mm} deduced from this diagram is consistent to that observed in Figure \ref{fig:snapG}.
We can also observe that the transmission coefficient is smaller for larger $\theta$ (also larger $\alpha$ for these cases as $\phi$ was kept constant), seen as the gap width between solutions widens at 
the equipartition layer ({ marked in the panel by a dot-dashed gray line} at $z=1.3$ Mm). This is consistent to 
the result obtained by \cite{Paul1}, and to the comparison of  the previous simulations G-s and G-sr.
 
Right panel of Figure \ref{fig:p2} shows that for larger $\theta$ the damping of the amplitude of the vertical velocity appears earlier in the atmosphere for larger $\theta$ 
for the fast wave solution (solution 3). This is consistent with the results obtained from the simulations, shown in Figure~\ref{fig:p1}.
%The points where the imaginary part of these solutions diverges in the ambipolar case compared to the MHD case are indicated with markers in the panel.
The ambipolar diffusion introduces a contribution to the electric field through a drift velocity which is
proportional to the Lorentz force. Thus, in the case of the fast waves, the velocity drift is proportional to the magnetic pressure gradient which is larger for propagation perpendicular to the field lines.

%%%%%%%%%%%%%%%%%%%%%%%%%%%%%%%%%%%%%%%%%%%%%%%%%%%%%%%%%2G
\begin{figure*}[!htb]
\centering
\FIG{\includegraphics[width=16cm]{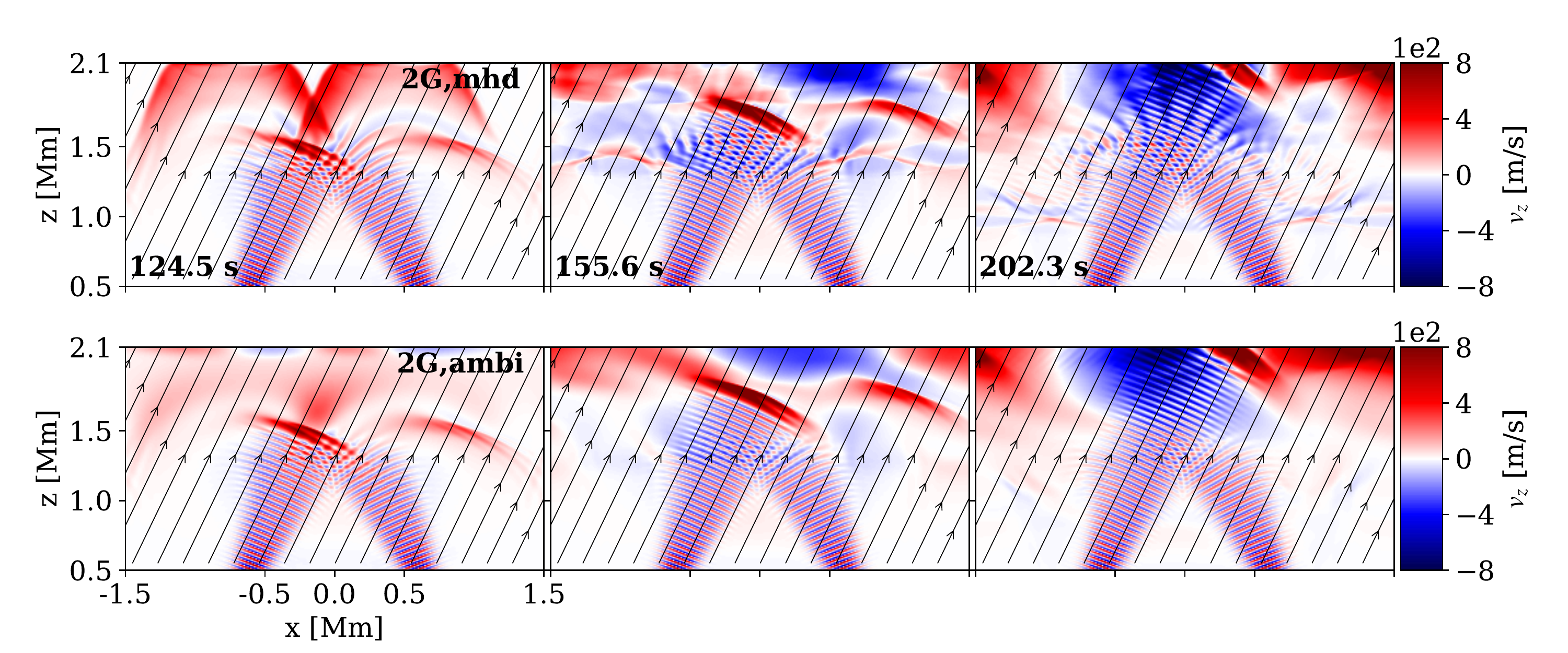}}
\caption{
Snapshots of the vertical velocity for the simulation 2G, the MHD case (upper panels) and ambipolar case (bottom panels)  taken at three moments of time.
}
\label{fig:snap2G_vz}
\end{figure*}
%%%%%%%%%%%%%%%%%%%%%%%%%%%%%%%%%%%%%%%%%%%
%%%%%%%%%%%%%%2G
%\begin{figure}[!htb]
%\centering
%\includegraphics[width=8cm]{snap_all_2G_velx.pdf}
%\caption{
%Snapshots of the horizontal velocity for the simulation 2G, the mhd case (upper panel) and ambipolar case (bottom panel)  taken at 198.1 s.
%}
%\label{fig:snap2G_vx}
%\end{figure}
%%%%%%%%%%%%%%%%%%%%%%%%%%%%%%%%%%%%%%%%%%%
\begin{figure*}[!htb]
\centering
\FIG{\includegraphics[width=16cm]{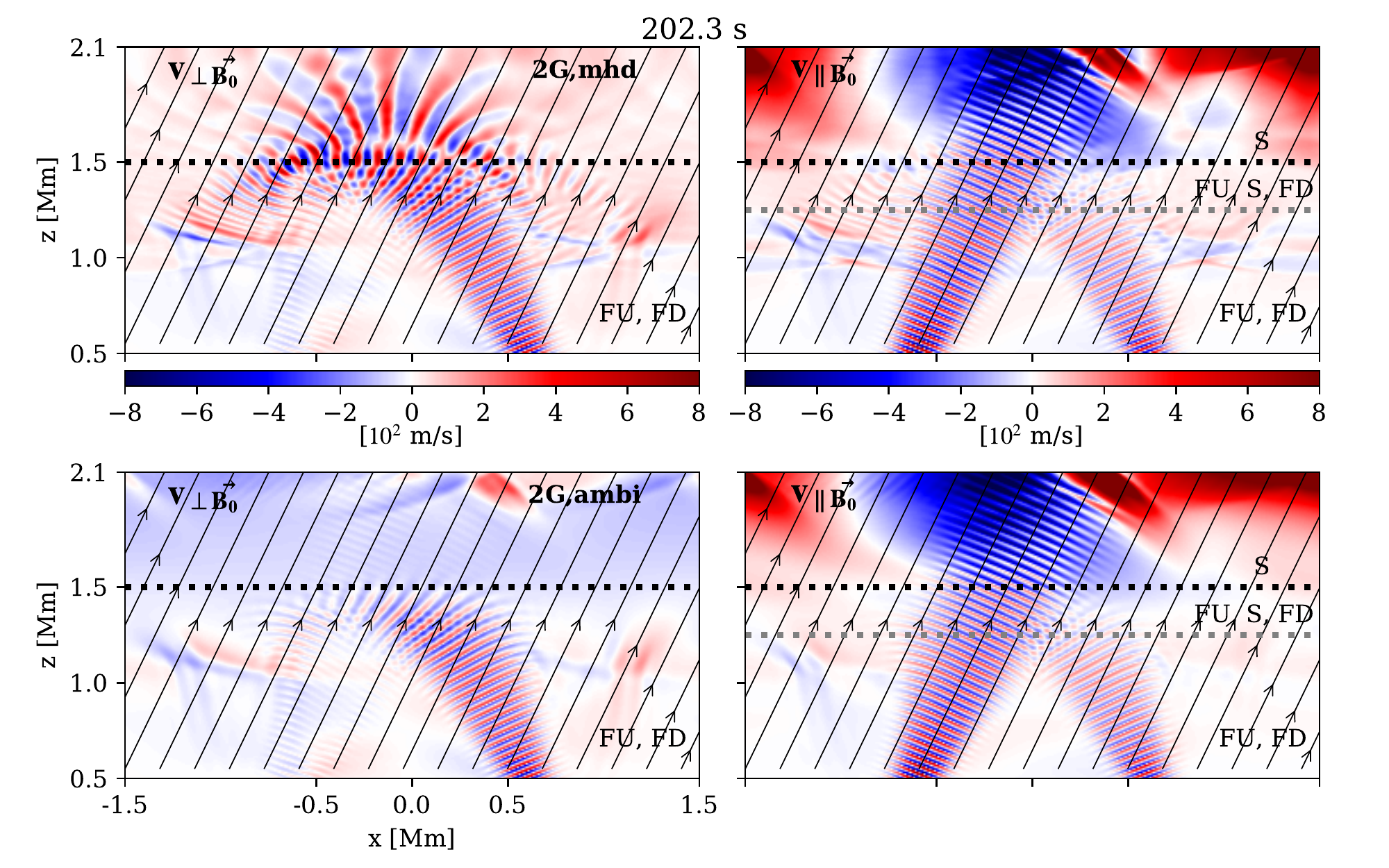}}
\caption{
Snapshots of the components of velocity perpendicular (left panels) and parallel (right panels) 
to the magnetic field for simulation 2G, for the MHD case (upper panels) and ambipolar case (bottom panels) taken at 198.1 s. 
This can be compared with the rightmost panels in Fig.~\ref{fig:snap2G_vz}, where we showed $v_z$ at the same time. 
The horizontal gray and black dotted lines present in the panels at the right hand side located at the heights $z=1.25$ Mm and $z=1.5$ Mm 
represent the equipartition layer and the reflection height, respectively. 
Above the reflection height the only wave which propagates is the slow mode, with the velocity oscillating along the magnetic field lines, observed in the right panel.
}
\label{fig:snap2G_vc}
\end{figure*}
%%%%%%%%%%%%%%%%%%%%%%%%%%%%%%%%%%%%%%%%%%%%
%\begin{figure*}[!htb]
%\centering
%\includegraphics[width=16cm]{snap_all_2G_mag.pdf}
%\caption{
%Snapshots of the vertical (left panels) and horizontal (right panels) componenets of the perturbation of 
%the magnetic field for the simulation 2G, the mhd case (upper panels) and ambipolar case (bottom panels) taken at 198.1 s.
%}
%\label{fig:snap2G_b1}
%\end{figure*}
%%%%%%%%%%%%%%%%%%%%%%%%%%%%%%%%%%%%%%%%%%%%%%%%%%
%
%%%%%%%%%%%%%%%%%%%%%%%%%%%%%%%%%%%%%%%%%%%%%%%%%%%%%%%%%%%%%%%%
\subsection{Varying the wave parameters}
%%%%%%%%%%%%%%%%%%%%%%%%%%%%%%%%%%%%%%%%%%%%%%%%%%%%%%%%%%%%%%%%
%
In order to observe the slow wave, we increase the angle of propagation at the bottom of the atmosphere to $\phi=49.5^\circ$, { compared to the P case}, 
so that the fast component reflects at a lower height. 
As the vertical wave number decreases, the solutions of  the local dispersion relation give a worse approximation.
We only show in
Figure \ref{fig:Pa} the profile of the vertical velocity at $x=0$, in the simulations P-a (left panel) and P-as (right panel), for both MHD and ambipolar runs. 
The angle of inclination  of the background magnetic field with the vertical direction for P-a is the same as in the simulation P, $\theta=25.8^\circ$.
For the P-as simulation the angle $\theta$ is increased, { this being the only difference to the P-a case.} 
{ For the P-as case, } $\theta=\phi$, enhancing the transmission of the slow wave.

In  both cases, in the lower part of the domain we can observe the fast upward wave driven at the bottom, before it interferes with the  reflected fast component.
The slow wave  can now be clearly seen as a wave packet after the 
reflection height of the fast wave.
We can observe that the fraction of the reflected component in the case P-as is larger than for the case P-s (see Figure \ref{fig:gs}, where there is no visible interference
because the wave is almost completely transmitted) where the direction of propagation at the base 
had also the same angle ($\phi$) as the background magnetic field  ($\theta$) with the vertical direction. The possible cause is the increase of the angle $\phi$.

We observe in Figure \ref{fig:Pa} that the fast components for both P-a and P-as reflect before the ambipolar term has any effect.
We can observe by comparing the two panels that the reflection height for P-a is larger than for P-as.
The damping of the slow  wave in the ambipolar case is more pronounced { at a lower height}.

We  then increase the plane wave period to $P=20$ s.
We consider the same angle $\phi=49.5^\circ$ and two values of $\theta$, namely $\theta=22.5^\circ$ (simulation P-p) and $\theta=84.2^\circ$ (simulation P-pf). 
{  The only difference between simulations P-p and  P-a presented above is a larger period for the P-p case.}

By using a local dispersion relation, \cite{ramp} find for waves which propagate in the vertical direction in an isothermal atmosphere, that there is a 
cutoff period  which increases with the density scale height. 
That means that waves with periods larger than the cutoff period do not propagate.
Their result was obtained for the slow modes in the general case, and for the fast modes in the case of horizontal magnetic field. 
For the slow modes, the cutoff period also increases with the inclination of the direction of the magnetic field.
In our case, as our temperature varies with height, this result applied locally means that 
waves with larger periods are reflected at smaller heights in the atmosphere.

The left and right panel of Figure \ref{fig:Pp} show a vertical cut at $x=0$ for the vertical velocity for the 20-second period cases P-p and P-pf, respectively. 
We can observe from the left panel of Figure \ref{fig:Pp} that the fast wave for P-p reflects at a smaller height compared to P-a,
{ which might be due to a larger period for the P-p case, this being the only difference between the two cases.}

From the right panel we can observe that the reflection height is increased for larger $\theta$.
{ This fact has been also observed for the case P-a compared to P-as, presented above.}
%This suggests that the results of \cite{ramp} might be valid in a more general context, including for the fast modes.
%%The ramp effect seems to be more efficient for larger periods, as it could not be observed for the period $P=5$ s,
%%considered before when  the waves reflect at the same height for different inclination angles, the cases P, G and P-f discussed above.
%
%The distance between the equipartition layer and the reflection height for the fast wave seems also to be smaller for the P-pf case,
%compared to the P-p case. This is suggested by the fact that we don't observe any interference region.
%The ambipolar damping of the slow component seems to be smaller for the P-p than for P-a, and the reason might be the larger period for P-p, this being the only
%difference between the two cases.

The fast component in the P-p case reflects before the ambipolar term has any effect. However, as the wave P-pf reflects at higher height we can observe the ambipolar effect on the fast component
above  $z\approx 1.25$ Mm for the P-pf case. The ambipolar effect { on the fast component} seems to be  smaller than for smaller periods, 
if we compare P-pf to P and G cases which have smaller $\alpha$ than P-pf.
{ The slow component  has a larger amplitude than the fast component  and is slightly more damped by the ambipolar diffusion for the P-p case compared to the P-a case.}

%
%%%%%%%%%%%%%%%%%%%%%%%%%%%%%%%%%%%%%%%%%%%%%%%%%%%%%%%%%%%
\subsection{Varying the magnetic field strength}
%%%%%%%%%%%%%%%%%%%%%%%%%%%%%%%%%%%%%%%%%%%%%%%%%%%%%%%%%%%
%
Figure \ref{fig:Pb} shows the vertical velocity profile  along $x=0$ for the simulation P-b at two moments of time.
All the parameters of the simulation P-b are the same as for the simulation P (from the left panel in Fig.~\ref{fig:p1}),
with the difference that the magnetic field is increased.
For this reason, the equipartition layer is now located at a lower height ($\approx$ 0.85 Mm) compared to the P case.

The panel at the left hand side shows a snapshot taken before the fast wave reaches the reflection height.
We can observe a similar interference 
between the fast and the slow wave as in the P simulation, with the difference that it starts at a smaller height.
We can also observe that the fast wave starts to be damped by the ambipolar diffusion earlier in the atmosphere, so that
the fast mode is more affected by the ambipolar diffusion in the P-b case as compared to the P case (see left panel of Figure \ref{fig:p1}).

The panel at the right hand side shows a snapshot taken before the wave reaches the upper boundary so that we can observe the slow transmitted wave.
The effect of the ambipolar diffusion on the slow wave seems to be larger for P-b compared to P-a (see left panel of Figure \ref{fig:Pa}), 
which can be attributed to a larger magnetic field, the angle $\alpha$ being larger 
in the P-a case, compared to the P-b case. The damping of the slow mode decreases towards the upper part of the atmosphere, similarly to P-a and P-as cases.

 Figure \ref{fig:dispB} compares the solutions of the dispersion relations for the case P and P-b.
Left panel shows the real part of  solutions 3 and 4, the upward fast and slow modes, respectively, for the MHD case for the two cases: P (black and gray solid lines)
and P-b (dark and light blue solid lines) as indicated in the legend.
The equipartition layer is indicated by vertical dot-dashed gray and light blue lines for P and P-b, respectively.
The reflection height of the fast component is indicated by a vertical dotted black line for P and a vertical dotted dark blue for P-b.

Right panel of Figure \ref{fig:dispB} shows the imaginary part of  solutions 3 and 4,  for the MHD case for the two cases: P  and P-b,
keeping the same color coding. Additionally we show the imaginary part of these solutions in the ambipolar case with dashed lines as indicated in the legend.
The points where the imaginary part of solution 3 in the MHD case diverges from the ambipolar case are indicated by a black and blue marker in the P and P-b case, respectively.
Above this diverging height the ambipolar effects produce damping of the fast wave components  compared to MHD cases.
We can observe that (1) the equipartition layer; (2) the diverging height; and (3) the reflection height; are all three shifted
to an earlier height in the P-b case compared to P. 

The slow modes are also more affected when the magnetic field is increased. This can be seen in the diagram 
as the area between the solution 4 for the ambipolar and MHD cases above the equipartition layer which is clearly larger for P-b
(area between pink dashed line and light blue solid line above $z\approx0.8$ Mm)  compared to P case (area between yellow dashed and gray solid lines above $z\approx1.25$ Mm).
However, the damping decreases in the upper part of the domain, where it seems that the ambipolar effects produce no damping of the slow modes.
This can be visually seen in the overlapping of the pink dashed line to the light blue solid line in the P-b case, and 
the yellow dashed to the  gray solid line in the P case.
These estimations are consistent with the results of the simulations.
\section{More complex setups: interacting pulses} \label{sec:nonlin}
%%%%%%%%%%%%%%%%%%%%%%%%%%%%%%%%%%%%%%%%%%%%%%%%%%%%%%%%%%%%%%%%%%%%%%%%%%%%%%%%%%%%%%%%%%%%%%%%%%%%%%%%%%%%%%%%%%%%%%
%%%%%%%%%%%%%%%%%%%%%%%%%%%%%%%%%%%%%%%%%%%%%%%%%%%%%%%%%%%%%%%%%%%%%%%%%%%%%%%%%%%%%%%%%%%%%%%%%%%%%%%%%%%%%%%%%%%%%%
%
We next consider a more complex case, where  
the perturbation used for simulation 2G is a superposition of the perturbations used in G-s and G-sr, where
the centers of the two gaussians are separated by 1.2 Mm. Hence, this 2G simulation involves two gaussians that will necessarily interact. The amplitude of the waves is also increased by a factor of 100, compared to the previous simulations,
 so that nonlinear effects become more important.
The interaction of the two waves creates small scales, an effect amplified by the fact that we used larger amplitudes for this simulation.
Figure \ref{fig:snap2G_vz} shows snapshots of vertical velocity  for the 2G simulation taken at different moments of time, and this for both the MHD and the ambipolar case.
%We can observe that the waves from the left gaussian pulse  which propagate almost along the field lines
%reflect when they reach the upper boundary, visually seen  in the negative flow right below the upper boundary (the dark blue region in the middle of the horizontal domain 
%at the location where the waves reach the upper boundary).
A small fraction of the left gaussian waves reflects at $z\approx1.5$ Mm, above the equipartition layer 
at the same height as the  waves from the right gaussian pulse. The right gaussian pulse is driven at the bottom of the atmosphere  
at a higher angle with respect to the magnetic field, and reflects almost completely. 
The behavior of the individual left and right gaussian pulses in the linear regime can be observed in Figure \ref{fig:gs}
in the snapshots of G-s and G-sr, respectively.

The fast driver wave is acoustic in nature in the deep layers where the magnetic field is weak, and the wave propagates isotropically.
%The  velocity oscillates mainly in the horizontal direction because of the gravity.
The gravity as a restoring force is less important than the gas pressure gradient, and the velocity oscillates along the direction of the propagation mainly.
Higher up in the atmosphere, above the equipartition layer located at $z\approx 1.25$ Mm, 
when the field becomes strong, the fast wave becomes magnetic in nature, and it propagates mainly across the field lines.
The slow wave there becomes acoustic in nature, and it propagates mostly along the field lines.
The dominant restoring force is the gradient of the gas pressure and the velocity oscillates mainly along the direction of propagation, parallel to the field lines.
The perturbation of the magnetic field is small for the slow modes.

Above the equipartition layer, where the dynamics is dominated by the magnetic field,
the fast and the slow wave can be highlighted by looking at the components of the velocity along and perpendicular to the magnetic field.
Figure \ref{fig:snap2G_vc} shows the components of the velocity perpendicular (left panels) and parallel (right panels) to the magnetic field, with the same colormap and normalization
as for Figure \ref{fig:snap2G_vz}, for an easier comparison. 
The top panels are for the MHD case, and the bottom panels are for the case when the ambipolar diffusion is present.
The horizontal black dotted lines located at $z=1.5$ Mm represent the
reflection height of the fast upward wave for the central component of the gaussian waves, which is a superposition of plane waves.
Additionally, the equipartition layer is shown in the right panels by the gray dotted line at $z=1.25$ Mm.
Below the reflection height (black dotted line) the fast upward and downward components can be seen in both left and right panel.

Above the equipartition layer, the fast upward waves change the direction of propagation to being perpendicular to the magnetic field lines, and this is better seen in the
left panels. Above the equipartition layer, the fast upward waves split into a fast component which reflects higher up (at the height about the black dotted line) and the slow component.
The velocity of the slow component oscillates mainly along the field lines so the slow component appears in the right panels. Above the reflection height, only the slow component
is present, seen in the right panels. The fact that we still see fast components a bit above the reflection height is related to the same argument mentioned before
for the gaussian perturbations, namely that different modes present in the superposition reflect at slightly different height.
The ambipolar effect, which starts to be important at about the equipartition layer, cannot be visually observed below this height.
Above the equipartition point, we can visually observe less damping in the ambipolar case in the right panel associated to the slow wave. 
%
%%%%%%%%%%%%%%%%%%%%%%%%%%%%%%%%%%%%%%%%%%%%%%%%%%%%%%%%%%%%%%%%%%%
\section{Analysis of wave heating}\label{sec:heat}
%%%%%%%%%%%%%%%%%%%%%%%%%%%%%%%%%%%%%%%%%%%%%%%%%%%%%
%\begin{figure*}[!htb]
%\centering
%\FIG{\includegraphics[width=8cm]{efub_ambi_pert2.pdf}
%\includegraphics[width=8cm]{meanEfUB_ambi_mhd_dif.pdf}}
%\includegraphics[width=8cm]{temp_z2.pdf}
%\caption{
%Left: Total energy vertical flux at the upper boundary for the mhd simulations. Right: Difference  in the total  energy vertical flux at the upper boundary.
%The vertical flux of the total energy has been averaged in the horizontal direction and over the last ten points of the vertical domain. 
%}
%\label{fig:efub}
%\end{figure*}
%%%%%%%%%%%%%%%%%%%%%%%%%
\begin{figure*}[!htb]
\centering
\FIG{
\includegraphics[width=8cm]{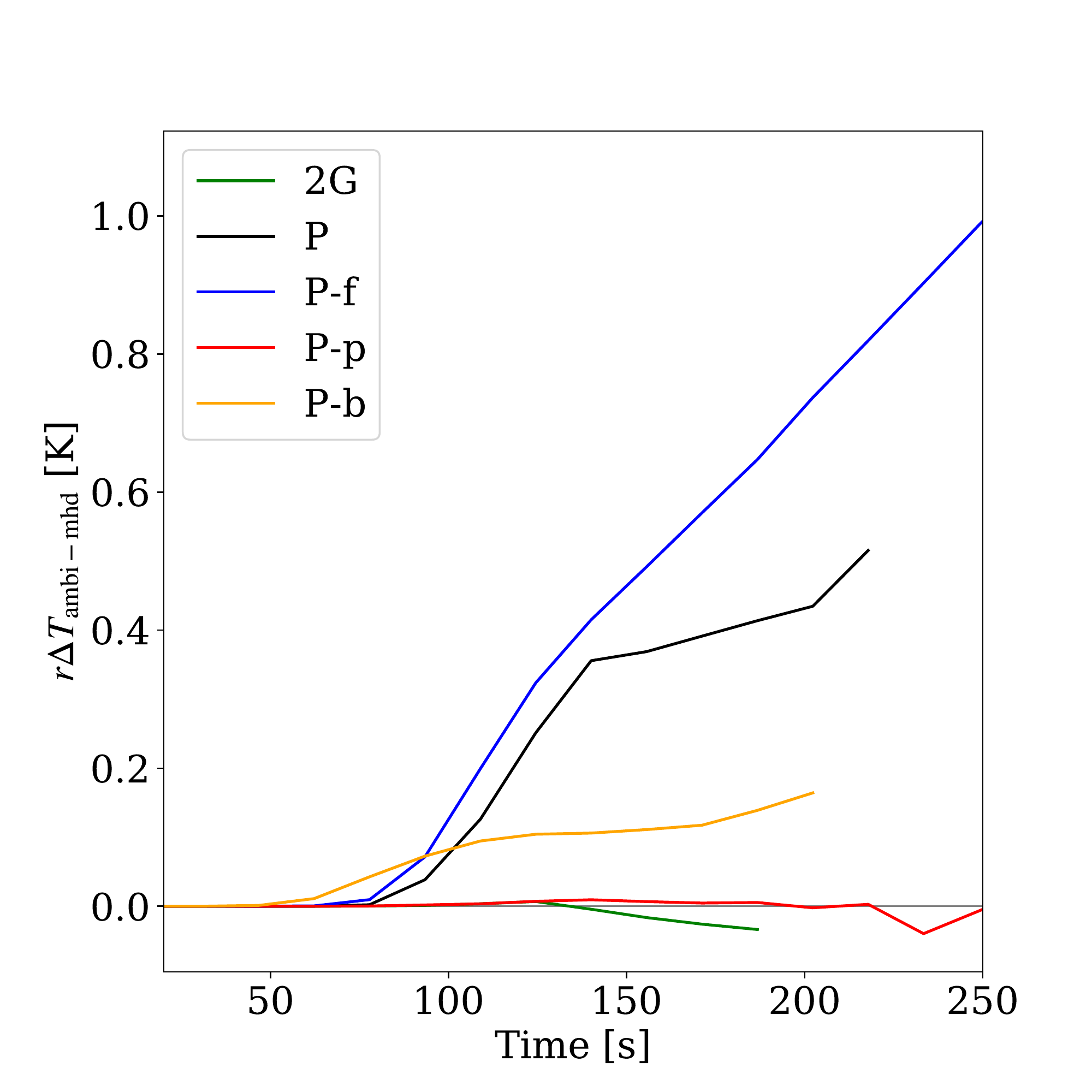}
\includegraphics[width=8cm]{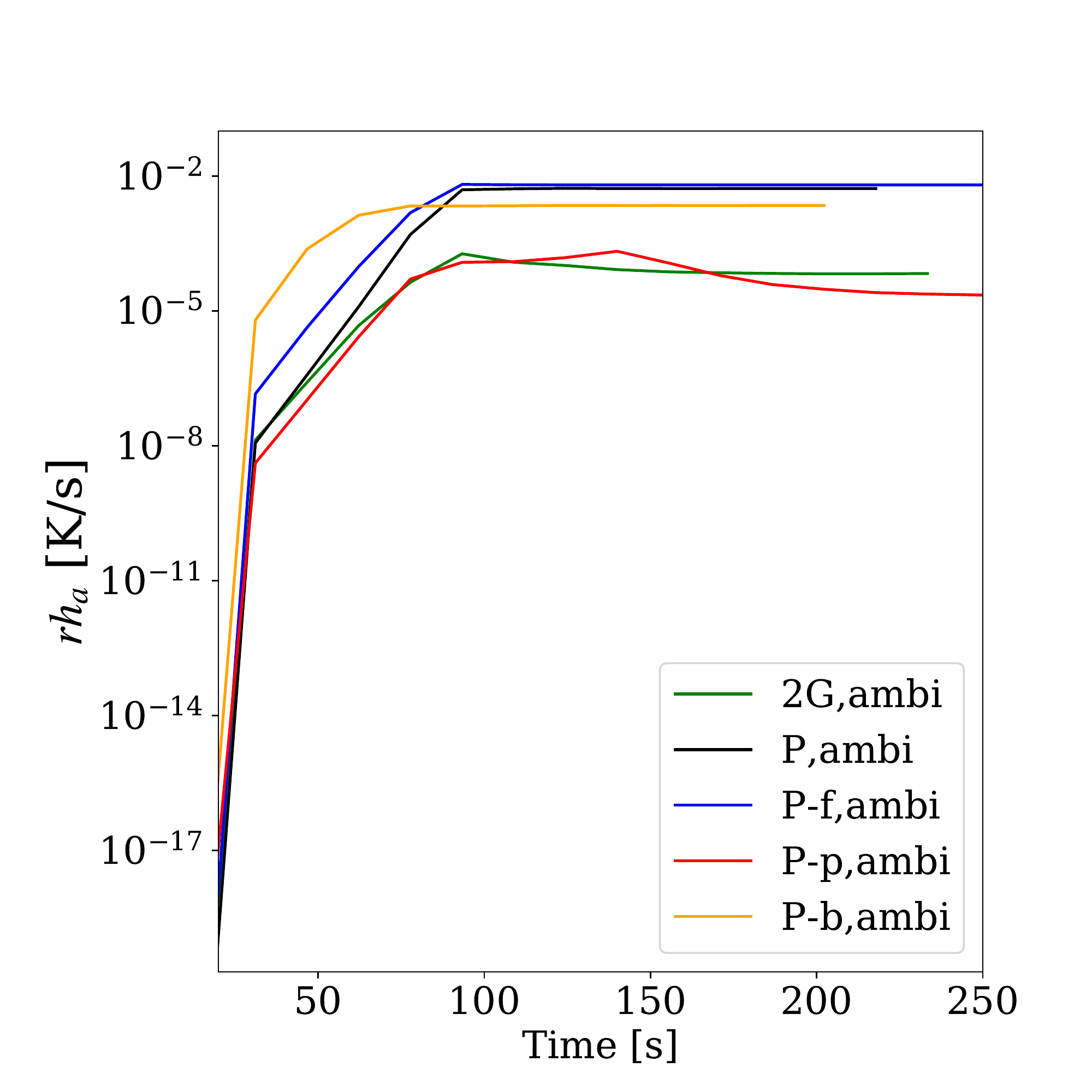}
}
\caption{
%Left: 
 {
Time evolution of the
average of the difference  in the mean temperature between the ambipolar and MHD simulations (left panel) and  the increase in temperature in a unit of time 
corresponding to the ambipolar heating $h_a$ (right panel).
The averages are done in the $z$-direction between the heights 0.8 and 1.7 Mm, and in the  $x$-direction.
For easier comparison, the quantities have been normalized using the value $r$, defined in Eq. (\ref{eq:rnorm}).}
%Right: Change in the mean magnetic energy compared to te equilibrium. {\bf RK You do not discuss the right panel, what is the point there? If not interesting, drop and keep only the left panel. In that left panel, do not yet include the ``2G,ambi,res2'' case.}
}
\label{fig:te1}
\end{figure*}
%%%%%%%%%%%%%%%%%%%%%%%%%%%%%%%%%%%%%%%%%%%%%%%%%%%%%
\begin{figure*}[!htb]
\centering
\FIG{\includegraphics[width=16cm]{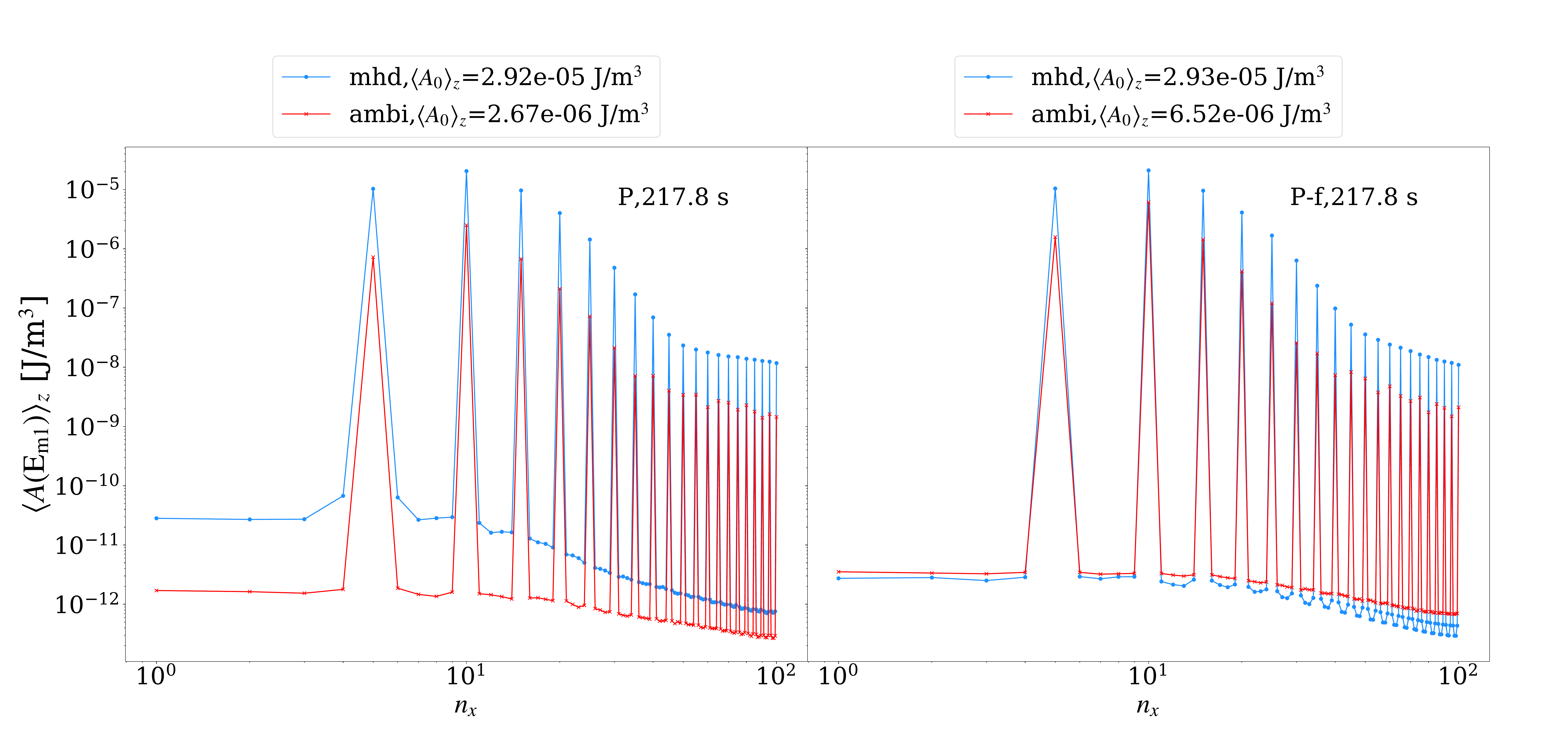}}
\caption{
Fourier amplitudes of the magnetic energy { perturbation} for the MHD simulations and the simulations with ambipolar effect included as a function of the horizontal  mode number
for the cases: P (left) and P-f (right). The amplitudes are calculated by a FFT transform in the $x$-direction for a snapshot taken at time  $t=217.8$ s and
averaged in height between 0.8 Mm and 1.7 Mm.
}
\label{fig:four}
\end{figure*}
%%%%%%%%%%%%%%%%%%%%%%%%%%%%%%%%%%%%%%%%%%%%%%%%%%%%%
%
%The ambipolar diffusion also adds a contribution to the Lorentz force. {\bf RK I do not understand the previous statement: there is always JxB, and this is identical to ideal MHD} 
 When the ambipolar diffusivity is present, the wave does work against the Lorentz force, and loses kinetic energy to magnetic energy.
The loss of the kinetic energy is seen in the damping of the wave in the ambipolar case. 
The ambipolar diffusion converts the  magnetic energy into heat through the dissipation of the electric field, resulting in an increase of temperature. 
The mean magnetic energy is smaller when  the ambipolar diffusion is considered, compared to pure MHD cases.

The increase in the temperature of the atmosphere depends on the total energy present in the domain, which is different for the various cases we showed, so it is difficult to compare between cases.
Energy is introduced at a constant rate by driving the wave through the bottom boundary, which also depends on the parameters of the wave (amplitude, wavevector, period).
At the open upper boundary, energy might escape or even enter the domain, and this variation in energy can depend on time.
These effects are supposedly similar for MHD versus ambipolar runs, so we will rather intercompare identical cases.
 
%As stated previously, our upper boundary treatment is not perfect (e.g. as seen and discussed in the net negative velocity profile from Fig.~\ref{fig:psd}). When we quantify the vertical flux of the total energy at our upper boundary as function of time, we find that e.g. the
%%In order to understand the evolution of the total energy in the domain we plot in Figure \ref{fig:efub} the total energy vertical flux at the upper boundary.
%%We averaged the flux in the horizontal direction and over the last ten points of the vertical domain.  
%strongly nonlinear 2G simulation has a large positive energy flux at our upper boundary, compared to the other simulations where this flux is overall negative 
%(ideally, this flux should be zero untill a wave arrives). 
%Also, for this 2G case, the top energy flux is larger for the ambipolar simulation compared to the MHD simulation. 

{
Left panel of Figure \ref{fig:te1} shows the difference in the average temperature  for several simulations in the MHD and ambipolar cases.
Right panel of Figure \ref{fig:te1} shows spatial average  of the ambipolar heating term, as a source term in the temperature equation,
\begin{eqnarray}\label{eq:heat}
h_a = \eta_A {J}_{\perp B}^2 B^2 \frac{\gamma-1}{\rho}  \,.
\end{eqnarray}
The spatial average has been done in the horizontal direction and in the vertical direction between heights $z=0.8$ and $z=1.7$ Mm.
For easier comparison between the 2G case, which used a larger amplitude and a different shape of the perturbation, with the rest of the cases,
the values have been multiplied by a normalization constant,
\begin{equation}\label{eq:rnorm}
r = \frac{1}{(l a)^2}\,,
\end{equation}
where $l$ is the fraction of the domain where the perturbation is applied, i.e. for the 2G case:
\begin{eqnarray}\label{eq:rnorm2G}
l=\frac{1}{3}\int_{-1.5}^{1.5}\Big[ \text{exp}\left(-\frac{(z-0.6)^2}{2 (0.08)^2}\right) + \nonumber \\  \text{exp}\left(-\frac{(z+0.6)^2}{2 (0.08)^2}\right) \Big]dz\,,
\end{eqnarray}
and $a$ is the amplitude ratio, having $a=100$ for the 2G case.
This gives the value of $r=5.625 \times 10^{-3}$ for the 2G case, and $r=1$ for all the other cases shown in Figures \ref{fig:te1}.
}

For all the simulations, except for the strongly nonlinear 2G case, the increase in temperature is larger in the ambipolar case when compared to its MHD equivalent. 
{ The conversion to heat is more efficient through compression than through the ambipolar heating in this case.}

{ The ambipolar heating term increases in time while the wave propagates towards the top of the domain, when it becomes flat.}
At later times, the formation of small scale structures enhances the ambipolar diffusion. The small scales are formed by the nonlinear interaction of 
the two gaussian pulses in the 2G case, or by the interaction of the  upwards going waves and the reflected fast downward waves for P, P-f and P-b cases.
In the P and P-b cases there are two waves propagating upwards above the equipartition layer, the slow and the fast modes, while for P-f only the fast mode exists.  
For this reason, the ambipolar heating for  the simulation P becomes similar to P-f simulation during late evolution, even if the latter has more magnetic energy.
In order to understand the distribution of the magnetic energy, 
we calculate next the Fourier spectra
in both simulations P and P-f for the MHD and ambipolar case.

Figure \ref{fig:four} shows the Fourier amplitudes of the magnetic energy { perturbation} (obtained using a fast Fourier transform in the $x$-direction) 
averaged in height between $z=0.8$ Mm and $z=1.7$ Mm
as a function of the horizontal mode number for the same time in the P  (left panel) and P-f (right panel) simulations. We show the spectra for both MHD (blue lines)
and ambipolar (red lines) cases. 
We can observe that the mean magnetic energy { perturbation} (the amplitude of mode $n=0$, shown in the legend) is larger for P-f simulation, compared to P simulation
for both MHD and ambipolar cases.
In the MHD case,  the amplitudes of the modes
with numbers which are multiples of the number of the driven mode ($n=5$)  are slightly larger for P-f compared to P,
however, intermediate scales (with mode numbers not multiples of 5), have more magnetic energy in P simulation compared to P-f.
The magnetic energy dissipated by ambipolar effect, which can be seen as the difference between the blue and the red curves,  seems to be larger for most of the scales in the P case,
especially for the smaller ones.
We conclude that  the existence of the additional slow transmitted mode  in the P simulation, 
besides  the fast upward and fast downward (reflected) modes present in both simulations, enhances 
 the ambipolar diffusion as a consequence of the formation of small scale structures through the (nonlinear) interaction of these waves.

%%%%%%%%%%%%%%%%%%%%%%%%%%%%%%%%%%%%%%%%%%%%%%%%%%%%%%%%%%%%%%%%%%%%%%%%%%%%%%%%%%%%%%%%%%%%%%%%%%%%%%%%%%%%%%%
\section{Conclusions} \label{sec:cl}

%%%TODO: SHow that the reflection from the boundary creates a wave with no structure in x

In this work we performed simulations of waves where we considered the ambipolar effect in the single fluid MHD model. Dispersion diagrams obtained from approximate analytical solutions to the linearized governing equations
helped us understand the results of the simulations. We varied parameters related to the background atmosphere and related to the properties of the waves, and studied how this impacts the effects of the ambipolar diffusion. 
The main conclusion is that ambipolar diffusion does not affect the wave propagation, but only the wave amplitude, damping the wave, similarly to the conclusions
obtained by \cite{wave2}. However, in this study we made a detailed analysis of the different parameters which can influence the results, with a clear separation between the effects on fast and slow modes.
As a secondary effect, wave interference structures formed from the interaction of different types of waves
are shown and seen to be differently affected by ambipolar diffusion. The main effects of the ambipolar diffusion are summarized below.

\begin{itemize}
\item Fast waves are more affected by ambipolar diffusion when their propagation is perpendicular to the magnetic field.
A larger  angle of propagation  between the direction of propagation at the base of the atmosphere and the magnetic field
makes the ambipolar damping effect larger also for the slow modes.

\item Ambipolar diffusion affects more the waves with smaller periods, for fast modes.
Fast waves with large angle of propagation with respect to the vertical direction, or large periods reflect before being affected by ambipolar diffusion.
A larger inclination of the magnetic field with the vertical direction
will allow fast components of waves with larger period to propagate higher up in the atmosphere where they can be damped by ambipolar diffusion.

\item 
The damping due to ambipolar diffusion is larger for larger magnetic field.
For the fast wave this is especially because the ambipolar damping starts earlier in the atmosphere. 
The slow modes are also more damped for larger magnetic field.

\item Slow waves are less affected by the ambipolar diffusion, compared to fast modes at the same height.
The damping of the slow waves decreases at larger heights.

\item { 
The mean temperature is generally larger for the simulations with ambipolar diffusion compared to the MHD simulations.
However, the compressive heating could be more efficient than the ambipolar heating.} 

\item
Small scale gradients enhance the ambipolar diffusion. In our simulations, the small scales are created by nonlinear effects, more pronounced
for larger amplitudes of the driven waves or by the (nonlinear) interaction of different modes.

\end{itemize}
%The turning pomts appears earleir in the atmosphere for the waves with larger period.

%It is then easy to see that there is not only enhanced diffusion, but there is also a contribution to the electromotive force proportional to the magnetic field \citep{turb1}

%$k_z$ in the deeep layers depend on the sound speed, the change in the magnetic fiels does not change the propagation speed.
%The movement accross the lines.
%
%The ambipolar diffusion does not change the topology of the lines
%For scales where the ambipolar effect is important,  
%the ambipolar effect seems to favor the mode conversion, there are slight changes in the real part, however the primary effect is the damoing which occurs earlier in the atmosphere.
%Asymptotically (for the fast waves) the height of the damping is the height where the real part changes.
%Without ambipolar the amplitude is larger and compensates for the enhanced mode transformation seen in the 
%
%The small scale structures created in all the cases by either the interference of the driven wave with the reflected wave 
%or by interefece between different driven waves will be affected by the ambipolar diffusion.

\begin{appendix}
\section{Comparing local dispersion relations for MHD}\label{appendixA}
The gravito-magneto-acoustic waves  described by two coupled second order ODEs, Eqs. (\ref{eq:mhd})  have exact analytical solution 
for an isothermal atmosphere in terms  of Meijer \citep{Zh1} { or hypergeometric \citep{Cally_2001}} functions.
However, this is not possible when the ambipolar diffusion is considered, moreover a local dispersion relation gives more intuitive insight about the behavior of the waves.

The assumption of a Fourier variation $\text{exp} (-ik_zz)$ in Eq.~(\ref{eq:mhd}) ignores the fact that the vertical direction is not uniform, but does lead to a local dispersion relation.
Doing so, we obtain a relationship between the amplitudes $V_x$, $V_z$ of the horizontal $v_x$ and vertical velocity $v_z$ perturbation, given by
\begin{align}\label{eq:mhd1}
&V_x = V_z \frac{N}{D}\,, \quad D\ne 0\,,\nonumber \\
&N = B_{\rm x0}  B_{\rm z0}  (k_x^2 + k_z^2)- \rho_0 c_0^2 k_x  k_z + i k_x \rho_0 g\,,\nonumber \\
&D = B_{\rm z0}^2  (k_x^2 + k_z^2) + \rho_0 \left(c_0^2   k_x^2 - \omega^2\right).
\end{align}
and a space dependent local dispersion relation, which is a fourth order equation in $k_z$, for a given real $\omega$ and $k_x$, hence of the form
\begin{equation}\label{eq:disp}
a_4 k_z^4 + a_3 k_z^3 + a_2 k_z^2 +  a_1 k_z +  a_0 = 0\,,
\end{equation}
where the coefficients are,
\begin{eqnarray}
a_4 = B_{\rm z0}^2 c_0^2 \rho_0 \,,\nonumber\\
a_3 = B_{\rm z0} \left[2 k_x B_{\rm x0} \gamma p_0- i B_{\rm z0} \gamma \rho_0 g\right]\,,\nonumber\\
a_2 = k_x^2 B_0^2 \gamma p_0 -i k_x B_{\rm x0} B_{\rm z0} \gamma \rho_0 g -\omega^2 \rho_0 (B_0^2+\gamma p_0) 
\,,\nonumber\\
a_1 = 2k_x^3 B_{\rm x0} B_{\rm z0} \gamma p_0+i(\omega^2\rho_0-k_x^2 B_{\rm z0}^2)\gamma\rho_0 g\,,\nonumber\\
a_0 = k_x^4 B_{\rm x0}^2\gamma p_0 -i k_x^3B_{\rm x0} B_{\rm z0} \gamma \rho_0 g \nonumber \\
-k_x^2\left[\rho_0^2g^2(1-\gamma)+\omega^2\rho_0(B_0^2+\gamma p_0)\right]+\omega^4\rho_0^2 \,.
\rule[0pt]{5pt}{0pt} 
\end{eqnarray}
The formal infinities which  appear in the atmosphere when $D=0$ in  the above Eq.~(\ref{eq:mhd1})  
do not concern us here, since our bottom driver avoids having this situation.

The WKB  method applied by \cite{Paul1} consists in first making a linear transformation of the variables so that 
the operator applied to the amplitudes of the new variables is self-adjoint.
This means that at zeroth order the amplitudes of the new variables are constant
and from a physical point of view, this is related to the conservation of energy
in a stratified atmosphere.
The combination used by \cite{Paul1} is given by:
\begin{equation}\label{eq:paulTransf}
\text{\boldmath $\xi$}\rightarrow \left(\rho_0^{\frac{1}{2}} c_0^2 \nabla \cdot \text{\boldmath $\xi$}, \rho_0^{\frac{1}{2}} a_0^2 (\xi_x \cos \theta -  \xi_z \sin \theta)  \right)\,,
\end{equation}
where $\text{\boldmath $\xi$} = (\xi_x,0,\xi_z)$ is the displacement variable, $a_0$ is the Alfv\'en speed, and $\theta$ is the angle of the magnetic field with the vertical direction. 
This $\text{\boldmath $\xi$}$ is related to the velocity through $\mathbf{v}=i\omega\text{\boldmath $\xi$}$. 
%
%H = rho0/d1rho0;
%omci2 = c02/(4*H^2);
%d1H = (d1rho0^2 - rho0 d2rho0)/d1rho0^2;
%(* omc2 = c02/(4*H^2) * (1-2 d1H); *)
%omc2 = c02/(4*H^2);
%K2 = kx^2 + kz^2;
%B2 = bz0^2 + bx0^2;
%a2 = B2/rho0;
%eps2 = G kx /om^2;
%
%N2= G/H - G^2/c02;
%
%kpar = kx bx0/Sqrt[B2] + kz bz0/Sqrt[B2];
%
%eqd = (1-eps2^2) om^4 - (a2 + (1-eps2^2)*c02 ) K2 om^2 + (1-eps2^2) a2 c02 K2 kpar^2 - ((1-eps2^2) om^2 - a2 K2)
%        (omc2 - (c02 N2 kx^2)/om^2)  - a2 K2 omci2 ((bx0^2/B2) + eps2^2 (bz0^2/B2)) + a2 c02 K2 kx eps2/H;
%
The dispersion relation obtained is Eq. (2.12) from \cite{Paul1}:
\begin{eqnarray}\label{dispPaul}
(1-\varepsilon^4) \omega^4 - \left[a_0^2 + (1-\varepsilon^4)c_0^2 \right] k^2 \omega^2 \nonumber\\ 
+ (1-\varepsilon^4) a_0^2 c_0^2 k^2 k_\parallel^2  \nonumber\\ 
     - \left[(1-\varepsilon^4) \omega^2 - a_0^2 k^2\right] \left[\omega_c^2 - (c_0^2 N^2 k_x^2)/\omega^2 \right]  \nonumber\\
    - a_0^2 k^2 \omega_{\rm ci}^2 \left[\text{sin}^2\theta + \varepsilon^4 \text{cos}^2\theta \right] %\nonumber\\
      + \frac{a_0^2 c_0^2 k^2 k_x \varepsilon^2}{H}=0\,, 
\end{eqnarray}
where:
\begin{eqnarray}\label{dispPaul2}
H=-\rho_0/(\frac{d \rho_0}{d z})\,,\quad
\omega_{\rm c}^2 = \omega_{\rm ci}^2 \left(1 - 2 \frac{d H}{d z}\right)\,,\nonumber\\ 
 \omega_{\rm ci}^2 = \frac{c_0^2}{4 H^2}\,,\quad N^2 = \frac{g}{H} - \frac{g^2}{c_0^2}\,,\quad
\varepsilon^2 = \frac{g k_x} {\omega^2}\,. \quad
\end{eqnarray}

When dissipation mechanisms are considered, the WKB method, as described by \cite{Paul1} cannot be applied as  no self-adjoint operator can be found.
In these cases, and even in dissipationless cases, when finding the right linear combination of variables was not straightforward,
the wave solution in the vertical direction was directly introduced in the linearized equations and local dispersion relations were obtained at the zeroth order which
neglects the vertical variations of the amplitudes.
The validity of the local dispersion relation is restricted to the range $|k_z| H \gg 1$, where $H$ is the density gradient scale length \citep{thomas2}.
The local dispersion relation is not unique and generally depends on the sequence of steps in the derivation \citep{thomas1}. 
\cite{thomas1,thomas2}  suggested that all the terms of order one or higher in $k_z H^{-1}$ should be neglected in order to obtain
a unique unambiguous form for the dispersion relation, further mentioned as approximate dispersion relation.
If we rather introduced the wave solution for the vertical direction directly in Eqs.~(\ref{eqs_mhd}), we would obtain a slightly different dispersion relation
than that described by Eq.~(\ref{eq:disp}),
where in this case with uniform magnetic field, the difference comes
only from the term $\partial p_1/\partial z$, which introduces terms related to equilibrium derivatives. 
For the same reason, changing variables to linear combinations of them which involve equilibrium profiles, changes slightly the dispersion relation obtained. 
%
%\begin{eqnarray}\label{stdCoef}
%%(bz0^2*c02)/(MU0*rho0)
%%Coef kz^3
%%(bz0*(I*bz0*G + 2*bx0*c02*kx))/(MU0*rho0)
%%Coef kz^2
%%(I*bx0*bz0*G*kx + bx0^2*(c02*kx^2 - om^2) + bz0^2*(c02*kx^2 - om^2) - c02*MU0*om^2*rho0)/(MU0*rho0)
%%Coef kz
%%(I*bz0^2*G*kx^2 + 2*bx0*bz0*c02*kx^3 - I*G*MU0*om^2*rho0)/(MU0*rho0)
%%Coef 0
%%(I*bx0*bz0*G*kx^3 + bx0^2*(c02*kx^4 - kx^2*om^2) + om^2*(-(bz0^2*kx^2) + MU0*(-(c02*kx^2) + om^2)*rho0))/(MU0*rho0)
%a_4 = B_{\rm z0}^2 c_0^2  \,,\nonumber\\
%a_3 = B_{\rm z0} \left[2 k_x B_{\rm x0} c_0^2 + i B_{\rm z0}  g\right]\,,\nonumber\\
%a_2 = (k_x^2 c_0^2 - \omega^2) B_0^2  + i k_x B_{\rm x0} B_{\rm z0}  g -\omega^2 \rho_0  c_0^2 
%\,,\nonumber\\
%a_1 = 2k_x^3 B_{\rm x0} B_{\rm z0} c_0^2 +i(-\omega^2\rho_0 + k_x^2 B_{\rm z0}^2) g\,,\nonumber\\
%a_0 = k_x^4 B_{\rm x0}^2c_0^2 -i k_x^3B_{\rm x0} B_{\rm z0}  g - k_x^2\omega^2(B_0^2+c_0^2 \rho_0)+\omega^4\rho_0 \,.
%\end{eqnarray}
%
The dispersion relation obtained  for an isothermal atmosphere, by directly introducing the vertical wave solution in the linearized equations 
is given by Eq. (3.6) from \cite{thomas2}, which for the 2D case is:
\begin{eqnarray}\label{dispThomas}
\omega^4 - (c_0^2 + v_{\rm A0}^2)  k^2  \omega^2 + g^2  (\gamma - 1)  k_x^2 \nonumber \\ 
+ \frac{(\mathbf{B_0}\cdot\mathbf{k})^2}{B_0^2}   c_0^2  v_{\rm A0}^2  k^2=0
\end{eqnarray}

%The local dispersion relation obtained at the zeroth order of the WKB method
%is generally different from the local  dispersion relation 
%in the standard form or any non-standard form, for the reason described by \cite{thomas1}.
%
%%%%%%%%%%%%%%%%%%%%%%%%%%%%%%%%%%%%%%%%%%%%%%%%%%%%%%%%%%%%%%%%%%%%%%%%%%%%%%%%%%%%%%%%%%%%%%%%%
\begin{figure*}[!htb]
\centering
\FIG{
\includegraphics[width=14cm]{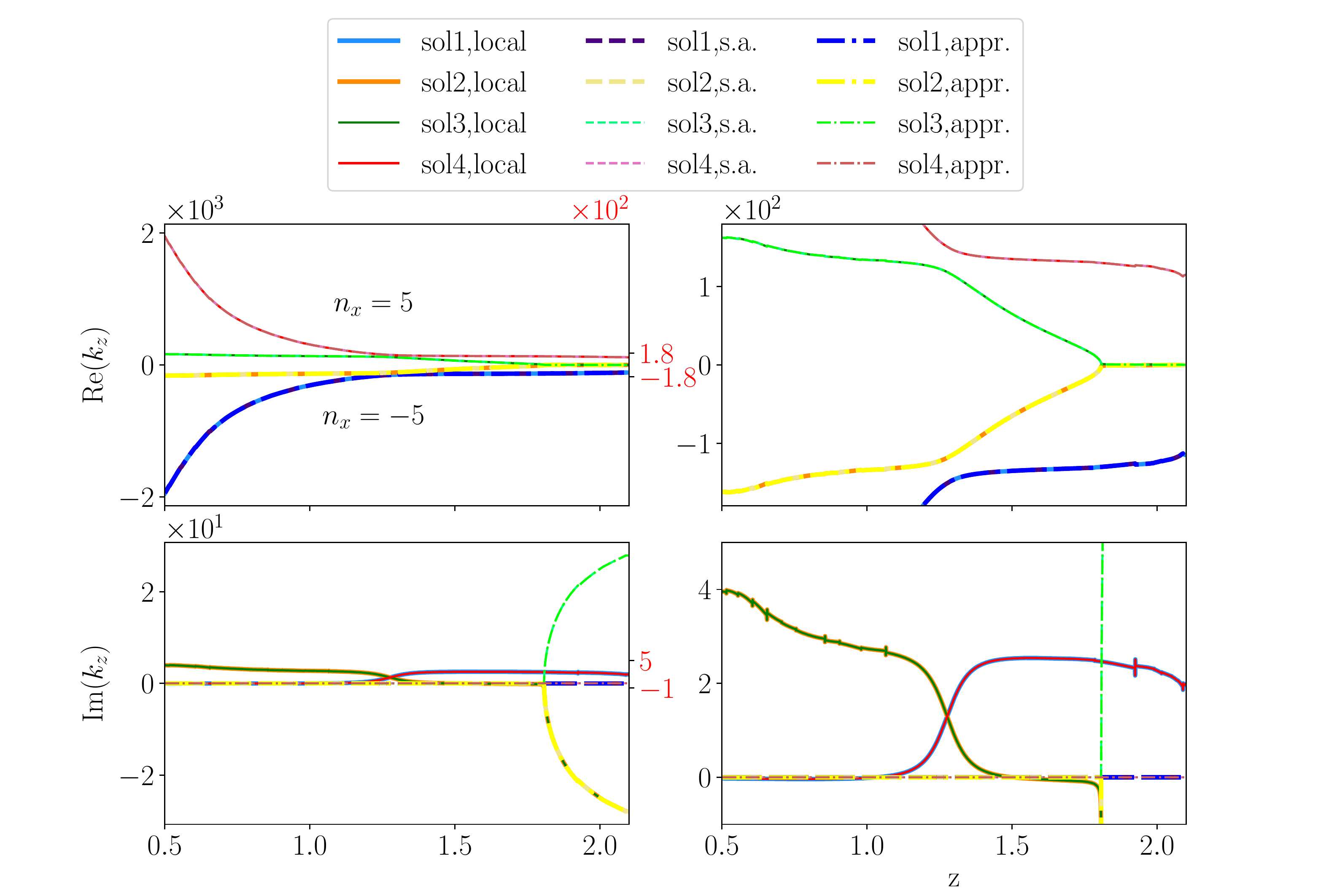}
}
\caption{
Case P.
The four solutions of the dispersion relation
for the MHD case in the three approximations: local dispersion relation, self-adjoint operator, approximate  dispersion relation.
Right panels show a zoom in the left panels for the limits shown as red ticks on the right axis of the left panels.
}
\label{fig:dispP}
\end{figure*}
%%%%%%%%%%%%%%%%%%%%%%%%%%%%%%%%%%%%%%%%%%%%%%%%%%%%%%%%%%%%%%%%%%%%%%%%%%%%%%%%%%%%%%%%%%%%%%%%%
%
Figure \ref{fig:dispP} compares the  four solutions of the three dispersion relations obtained in the MHD approximation: 
the dispersion relation used by \cite{Paul1} in the WKB approach, Eq.~(\ref{dispPaul}) (labeled ``s.a.''),
the  approximate dispersion relation obtained by \cite{thomas2}, i.e. Eq.~(\ref{dispThomas}) (labeled ``appr.'') and the local dispersion relation used in this work, Eq.~(\ref{eq:disp}) (labeled ``local'').
The dispersion relation has been solved for each height using the local parameters of the background atmosphere, corresponding to the simulation P.
The right panels correspond to a zoom indicated on the right axis of the left panels.
The downward propagating solutions have been obtained by reversing the horizontal wave number, indicated by the labels ``$n_x=5$'' and ``$n_x=-5$''.
In this way we can test  that the upward and downward solutions for the real part are symmetric.
The real part is related to  wave propagation, solutions 1 and 4 correspond to the slow waves which propagate downwards and upwards respectively and solution 2 and 3 correspond to
downward and upward fast waves.
The imaginary part is related to the amplification/damping of the amplitude,
for the waves propagating upwards a positive value being related to amplification and a negative value to damping, and the reverse for the downward waves.

We can observe that the three approximations give similar symmetric solutions for the real part.
The imaginary part of the solutions  are however different in the three approximations.
The dispersion relation obtained from a self-adjoint operator, ``s.a.'' has the imaginary or real  part exactly zero, as expected.
In this case the amplitude of the wave, at zeroth order should be calculated by the constraint that 
the amplitudes of the two transformed variables, described by Eq.~(\ref{eq:paulTransf}) are constant.
The ``local'' dispersion relation solutions  2 and 3, as well as 1 and 4 have the same imaginary part, suggesting the space reversibility, mathematically described by a possible self-adjoint formulation.
This does not happen when the ambipolar effect is taken into account, as the damping due to the dissipation effects appears in  both upwards and downwards propagation.
The ``local'' dispersion relation  has a positive imaginary part for the fast upward wave (solution 3) at the bottom of the atmosphere, up above the equipartition layer (where the curves 
of the real part for the solution 3 and 4 approach).
This is consistent with the fact that the amplitude of the vertical  velocity of the fast wave, acoustic in this region, grows, while it propagates upwards and the density drops.
The imaginary part for the approximate dispersion relation (``approx'') is zero for all the solutions in the whole domain
except for the fast components above their reflection height, in particular for the fast upward wave in the lower part of the atmosphere,
being unable to capture the stratification. 
%We checked that this was not due to the isothermal assumption, the dispersion relation, constructed
%using the same recipe as suggested by \cite{thomas2}, without this assumption did not gave better results (not shown here).

The higher order correction in the WKB approach, which describes the space dependence of the amplitude cannot be obtained analytically, in the general case, 
for a system of two coupled ODEs without further assumptions,
other than only neglecting the second order derivatives of the amplitudes, as in the original WKB approach.
However, by neglecting both second order derivatives and products of first order derivatives in the amplitude and the vertical wavenumber, a first order correction can be obtained
in a straightforward way, when 
the gradient in the amplitude can be calculated from the gradient in the vertical wave number, which can be introduced as a correction for the wavenumber,
as shown in \cite{Popescu2019}.
This correction was negligible for this case, meaning that the ``local'' dispersion relation gives good results for this level of approximation.
%On the other hand, both the dispersion relation and the more general WKB method can give erroneous results, compared to the exact solution \citep{thomas2}.
%There is no known exact analytical solution for the equations which take into account the  ambipolar diffusion in the general case.
%When the ambipolar effect is considered in the 2D propagation of waves in an atmosphere with an oblique magnetic field, the equations complicate considerably
%and and there is no straightforward approach. 
The comparison between the results from the simulation to the ``local'' dispersion relation gave  good agreement.
The analysis presented in this Appendix suggests that the ``local'' dispersion relation used in this work was our best choice for this study.
\end{appendix}
\begin{acknowledgements}
This work was supported by the ERC Advanced Grant PROMINENT and a joint FWO-NSFC grant G0E9619N. This
project has received funding from the European Research Council (ERC) under
the European Union’s Horizon 2020 research and innovation programme (grant
agreement No. 833251 PROMINENT ERC-ADG 2018). This research is further supported by Internal funds KU Leuven, through a PDM mandate and the project C14/19/089 TRACESpace.
\end{acknowledgements}

%\clearpage
\bibliographystyle{aa}
%\bibliography{biblio}

\begin{thebibliography}{37}
\expandafter\ifx\csname natexlab\endcsname\relax\def\natexlab#1{#1}\fi

\bibitem[{Alexiades {et~al.}(1996)Alexiades, Amiez, \& Gremaud}]{sts2-ref}
Alexiades, V., Amiez, G., \& Gremaud, P.-A. 1996, Communications in Numerical
  Methods in Engineering, 12, 31

\bibitem[{Bai \& Stone(2011)}]{mri}
Bai, X.-N. \& Stone, J.~M. 2011, The Astrophysical Journal, 736, 144

\bibitem[{{Bel} \& {Leroy}(1977)}]{ramp}
{Bel}, N. \& {Leroy}, B. 1977, \aap, 55, 239

\bibitem[{{Brandenburg}(2019)}]{turb1}
{Brandenburg}, A. 2019, \mnras, 487, 2673

\bibitem[{Cally(2001)}]{Cally_2001}
Cally, P.~S. 2001, The Astrophysical Journal, 548, 473

\bibitem[{{Cally}(2005)}]{Paul2}
{Cally}, P.~S. 2005, \mnras, 358, 353

\bibitem[{Cally(2006)}]{Paul1}
Cally, P.~S. 2006, Philosophical Transactions: Mathematical, Physical and
  Engineering Sciences, 364, 333

\bibitem[{{Cally} \& {Khomenko}(2018)}]{wave2}
{Cally}, P.~S. \& {Khomenko}, E. 2018, \apj, 856, 20

\bibitem[{Goedbloed {et~al.}(2019)Goedbloed, Keppens, \& Poedts}]{hans2019}
Goedbloed, H., Keppens, R., \& Poedts, S. 2019, Magnetohydrodynamics of
  Laboratory and Astrophysical Plasmas (Cambridge University Press)

\bibitem[{{Gonz{\'a}lez-Morales} {et~al.}(2018){Gonz{\'a}lez-Morales},
  {Khomenko}, {Downes}, \& {de Vicente}}]{sts2-ref2}
{Gonz{\'a}lez-Morales}, P.~A., {Khomenko}, E., {Downes}, T.~P., \& {de
  Vicente}, A. 2018, \aap, 615, A67

\bibitem[{Hoyos {et~al.}(2010)Hoyos, Reisenegger, \& Valdivia}]{i2}
Hoyos, J.~H., Reisenegger, A., \& Valdivia, J.~A. 2010, Monthly Notices of the
  Royal Astronomical Society, 408, 1730

\bibitem[{{Jones}(1987)}]{i1}
{Jones}, P.~B. 1987, \mnras, 228, 513

\bibitem[{Keppens {et~al.}(2021)Keppens, Teunissen, Xia, \& Porth}]{amrvac}
Keppens, R., Teunissen, J., Xia, C., \& Porth, O. 2021, Computers \&
  Mathematics with Applications, 81, 316, development and Application of
  Open-source Software for Problems with Numerical PDEs

\bibitem[{{Khomenko}(2009)}]{khrev}
{Khomenko}, E. 2009, in Astronomical Society of the Pacific Conference Series,
  Vol. 416, Solar-Stellar Dynamos as Revealed by Helio- and Asteroseismology:
  GONG 2008/SOHO 21, ed. M.~{Dikpati}, T.~{Arentoft}, I.~{Gonz{\'a}lez
  Hern{\'a}ndez}, C.~{Lindsey}, \& F.~{Hill}, 31

\bibitem[{{Khomenko} {et~al.}(2014){Khomenko}, {Collados}, {D{\'\i}az}, \&
  {Vitas}}]{eqkh}
{Khomenko}, E., {Collados}, M., {D{\'\i}az}, A., \& {Vitas}, N. 2014, Physics
  of Plasmas, 21, 092901

\bibitem[{{Meyer} {et~al.}(2014){Meyer}, {Balsara}, \& {Aslam}}]{sts1-ref1}
{Meyer}, C.~D., {Balsara}, D.~S., \& {Aslam}, T.~D. 2014, Journal of
  Computational Physics, 257, 594

\bibitem[{Meyer {et~al.}(2014)Meyer, Balsara, \& Aslam}]{sts1-ref}
Meyer, C.~D., Balsara, D.~S., \& Aslam, T.~D. 2014, Journal of Computational
  Physics, 257, 594

\bibitem[{{Musielak} {et~al.}(1994){Musielak}, {Rosner}, {Stein}, \&
  {Ulmschneider}}]{mus}
{Musielak}, Z.~E., {Rosner}, R., {Stein}, R.~F., \& {Ulmschneider}, P. 1994,
  \apj, 423, 474

\bibitem[{{N{\'o}brega-Siverio} {et~al.}(2020){N{\'o}brega-Siverio},
  {Mart{\'\i}nez-Sykora}, {Moreno-Insertis}, \& {Carlsson}}]{sts2-ref1}
{N{\'o}brega-Siverio}, D., {Mart{\'\i}nez-Sykora}, J., {Moreno-Insertis}, F.,
  \& {Carlsson}, M. 2020, \aap, 638, A79

\bibitem[{{Nye} \& {Thomas}(1976)}]{Nye}
{Nye}, A.~H. \& {Thomas}, J.~H. 1976, \apj, 204, 573

\bibitem[{{O'Sullivan} \& {Downes}(2006)}]{sts2-ref3}
{O'Sullivan}, S. \& {Downes}, T.~P. 2006, \mnras, 366, 1329

\bibitem[{O'Sullivan \& Downes(2007)}]{sts2-ref4}
O'Sullivan, S. \& Downes, T.~P. 2007, Monthly Notices of the Royal Astronomical
  Society, 376, 1648

\bibitem[{{Popescu Braileanu}(2020)}]{mythesis}
{Popescu Braileanu}, B. 2020, PhD thesis, University of La Laguna,
  https://www.educacion.es/teseo/mostrarRef.do?ref=1866768

\bibitem[{{Popescu Braileanu} {et~al.}(2019){Popescu Braileanu}, {Lukin},
  {Khomenko}, \& {de Vicente}}]{Popescu2019}
{Popescu Braileanu}, B., {Lukin}, V.~S., {Khomenko}, E., \& {de Vicente},
  {\'A}. 2019, \aap, 630, A79

\bibitem[{{Schunker} \& {Cally}(2006{\natexlab{a}})}]{wave1}
{Schunker}, H. \& {Cally}, P. 2006{\natexlab{a}}, in ESA Special Publication,
  Vol. 624, Proceedings of SOHO 18/GONG 2006/HELAS I, Beyond the spherical Sun,
  ed. K.~{Fletcher} \& M.~{Thompson}, 5

\bibitem[{{Schunker} \& {Cally}(2006{\natexlab{b}})}]{Paul1new}
{Schunker}, H. \& {Cally}, P.~S. 2006{\natexlab{b}}, \mnras, 372, 551

\bibitem[{{Sinha}(2020)}]{i3}
{Sinha}, M. 2020, arXiv e-prints, arXiv:2010.10776

\bibitem[{{Spruit} \& {Bogdan}(1992)}]{bog1}
{Spruit}, H.~C. \& {Bogdan}, T.~J. 1992, \apjl, 391, L109

\bibitem[{{Stone} {et~al.}(2020){Stone}, {Tomida}, {White}, \&
  {Felker}}]{sts1-ref3}
{Stone}, J.~M., {Tomida}, K., {White}, C.~J., \& {Felker}, K.~G. 2020, \apjs,
  249, 4

\bibitem[{{Thomas}(1982)}]{thomas1}
{Thomas}, J.~H. 1982, \apj, 262, 760

\bibitem[{{Thomas}(1983)}]{thomas2}
{Thomas}, J.~H. 1983, Annual Review of Fluid Mechanics, 15, 321

\bibitem[{Toro(1997)}]{hll}
Toro, E.~F. 1997, The HLL and HLLC Riemann Solvers (Berlin, Heidelberg:
  Springer Berlin Heidelberg), 293--311

\bibitem[{{Vernazza} {et~al.}(1981){Vernazza}, {Avrett}, \& {Loeser}}]{VALC}
{Vernazza}, J.~E., {Avrett}, E.~H., \& {Loeser}, R. 1981, \apjs, 45, 635

\bibitem[{{Xia} \& {Keppens}(2016)}]{sts1-ref2}
{Xia}, C. \& {Keppens}, R. 2016, \apj, 823, 22

\bibitem[{{Xia} {et~al.}(2017){Xia}, {Keppens}, \& {Fang}}]{sts1-ref4}
{Xia}, C., {Keppens}, R., \& {Fang}, X. 2017, \aap, 603, A42

\bibitem[{{Zhugzhda} \& {Dzhalilov}(1984)}]{Zh1}
{Zhugzhda}, I.~D. \& {Dzhalilov}, N.~S. 1984, \aap, 132, 45

\bibitem[{Čada \& Torrilhon(2009)}]{cada3}
Čada, M. \& Torrilhon, M. 2009, Journal of Computational Physics, 228, 4118

\end{thebibliography}

\end{document}